\documentclass[11pt,a4paper]{amsart}

\usepackage[latin9]{inputenc}
\usepackage{amsthm}
\usepackage{amstext}
\usepackage{amssymb}
\usepackage{caption}

\makeatletter

\pdfpageheight\paperheight
\pdfpagewidth\paperwidth

\numberwithin{equation}{section}
\numberwithin{figure}{section}

\usepackage{amsthm}
\usepackage{graphicx}
\usepackage{txfonts}
\usepackage{pxfonts}
\usepackage{marginnote}

\pdfpageheight\paperheight
\pdfpagewidth\paperwidth

\numberwithin{equation}{section}
\numberwithin{figure}{section}

\usepackage{amsthm}
\usepackage{amsthm}
\usepackage{epic}
\usepackage{eepic}
\usepackage{float}
\usepackage{rotating}
\usepackage{epsfig}
\usepackage{indentfirst}
\usepackage{array}
\usepackage{varioref}
\usepackage{tikz}
\usepackage{marginnote}
\usepackage{pgf}
\usepackage{bbm}
\usepackage{appendix}

\usepackage{color}

\let\emptyset\varnothing

\def\ll{\left\lgroup}
\def\rr{\right\rgroup}

  \def\rrho{ {\tt \scriptstyle R} }
\def\ssigma{ {\tt \scriptstyle C} }
    \def\MM{ {    \scriptstyle M} }

\def\leq{\leqslant}
\def\geq{\geqslant}

\def\pp{p^{\prime}}

\restylefloat{figure}

\newcommand{\cA}{\mathcal{A}}
\newcommand{\cB}{\mathcal{B}}
\newcommand{\cC}{\mathcal{C}}

\newcommand{\cF}{\mathcal{F}}

\newcommand{\cH}{\mathcal{H}}

\newcommand{\cL}{\mathcal{L}}
\newcommand{\cM}{\mathcal{M}}
\newcommand{\cN}{\mathcal{N}}
\newcommand{\cO}{\mathcal{O}}

\newcommand{\cV}{\mathcal{V}}
\newcommand{\cW}{\mathcal{W}}
\newcommand{\cX}{\mathcal{X}}

\newcommand{\cZ}{\mathcal{Z}}

\newcommand{\QQ}{\mathbbm{Q}}

\newcommand{\ZZ}{\mathbbm{Z}}

\newcommand{\apos}{\alpha_{+}}
\newcommand{\aneg}{\alpha_{-}}

\newcommand{\ws}{\square}
\newcommand{\bs}{\blacksquare}
\newcommand{\eo}{\epsilon_1}
\newcommand{\et}{\epsilon_2}

\def\ll{ \left\lgroup}
\def\rr{\right\rgroup}

\begin{document}

\title[]{From topological strings to minimal models}

\author[]{Omar Foda \!$^{{\scriptstyle \mathbbm{\, A      }}}$ and 
       Jian-Feng Wu \!$^{{\scriptstyle \mathbbm{\, B, \, C}}}$}

\address{
\!\!\!\!\!\!\!\!\!
$^{{\scriptstyle \mathbbm{A}}}$ 
School of Mathematics and Statistics, University of Melbourne, 
Royal Parade, Parkville, VIC 3010, Australia
\newline
$^{{\scriptstyle \mathbbm{B}}}$ 
Department of Mathematics and Statistics, Henan University, 
Minglun Street, Kaifeng city, Henan, China
\newline
$^{{\scriptstyle \mathbbm{C}}}$ 
Institute of Theoretical Physics and Mathematics, 
3rd Shangdi Street, Beijing, China}

\email{omar.foda@unimelb.edu.au, muchen.wu@gmail.com}


\begin{abstract}
We glue four refined topological vertices to obtain the building block of 5D $U(2)$ 
quiver instanton partition functions. We take the 4D limit of the result to obtain 
the building block of 4D instanton partition functions which, using the AGT correspondence, 
are identified with Virasoro conformal blocks.

We show that there is a choice of the parameters of the topological vertices that we start 
with, as well as the parameters and the intermediate states involved in the gluing procedure, 
such that we obtain Virasoro minimal model conformal blocks. 
\end{abstract}

\maketitle


\section{Introduction}
\label{section.01.introduction}
{\it 
We plan to start from the refined topological vertex and obtain Virasoro $A$-series minimal model 
conformal blocks times Heisenberg factors.
}

\medskip

\subsection{Background and basic concepts}

\subsubsection{Topological strings} Perturbative string theory, seen from a world-sheet point of view, 
is a 2D conformal field theory, coupled to 2D gravity, on Riemann surfaces. If the 2D conformal field 
theory is topological, in the sense that the correlation functions are independent of the metrics on 
the Riemann surfaces, the resulting string theory is {\it topological}. For an introduction, we refer 
to \cite{vafa.neitzke}, and references therein.
There are two constructions of topological string theories, the A-model and the B-model. For the 
purposes of this note, it suffices to say that we work are in the context of the A-model, and that 
the target space, that the strings propagate in, is $\mathbb{R}^{\, 1, 3} \times \cX$, where $\cX$ 
is a toric Calabi-Yau complex 3-manifold. For suitable choices of $\cX$, topological strings are 
non-trivial but tractable, and one can compute their partition functions. For further details, we 
refer to the reviews \cite{marino.2005, marino.book}, and references therein.

\subsubsection{From topological strings to 5D gauge theories} Topological string partition functions 
are interesting in themselves, and for yet another, an {\it apriori} unexpected reason. Namely, for 
suitable choices of the Calabi-Yau manifold $\cX$, one can compute the topological string partition 
function $\cZ^{\, top}$, and moreover, identify the result with the instanton partition function 
$\cZ^{\, 5D}_{instanton}$ of a corresponding 5D $\cN \! = \! 2$ supersymmetric quiver gauge theory,
thereby {\it \lq geometrically-engineering\rq} the latter theory 
\cite{geometric.engineering.01, geometric.engineering.02}. 

\subsubsection{Refined partition functions} A 5D instanton partition function depends, in general, 
on two deformation parameters, $\epsilon_1$ and $\epsilon_2$. 
For $\epsilon_1 + \epsilon_2 = 0$, the instanton partition function is unrefined. 
For $\epsilon_1 + \epsilon_2 \neq 0$, the instanton partition function is refined. 
Given the identification of 5D 
instanton partition function and topological strings, the latter also depend, in general on 
the deformation parameters,  $\epsilon_1$ and $\epsilon_2$.
For $\epsilon_1 + \epsilon_2 = 0$, the topological string partition function is unrefined.
For $\epsilon_1 + \epsilon_2 \neq 0$, the instanton partition function is refined.  

\subsubsection{The refined topological vertex} To compute topological string partition functions,
one splits the full partition function into basic building blocks, computes the contribution of each 
building block, then combines the contributions to obtain the required result.
In the unrefined case, $\epsilon_1 + \epsilon_2 = 0$, this was achieved in \cite{topological.vertex}. 
In this case, the building blocks are copies of the original, unrefined {\it \lq topological vertex\rq} 
introduced in \cite{topological.vertex}.
In the   refined case, $\epsilon_1 + \epsilon_2 \neq 0$, this was achieved in 
\cite{awata.kanno.01, awata.kanno.02, ikv}
In this case, the building blocks are copies of the {\it \lq refined topological vertex\rq} 
\cite{awata.kanno.01, awata.kanno.02, ikv}. 

\subsubsection{From 5D to 4D gauge theories} 
A 5D quiver gauge theory instanton partition function, $\cZ^{\, 5D}_{instanton}$, depends on the radius 
$R$ of a space-like circle. In the limit $R \rightarrow 0$, $\cZ^{\, 5D}_{instanton}$ reduces to 
a corresponding 4D instanton partition function $\cZ^{\, 4D}_{instanton}$ 
\cite{nekrasov, hollowood.iqbal.vafa, ikv, eguchi.kanno, zhou, taki.01}.
The deformation parameters $\epsilon_1$ and $\epsilon_2$ of the 5D theory are inherited by the 4D theory.

\subsubsection{From 4D gauge theories to generic 2D conformal field theories} The 4D instanton partition 
functions $\cZ^{\, 4D}_{instanton}$ are identified {\it via} the AGT correspondence with 2D Virasoro 
generic conformal blocks times Heisenberg factors, $\cB^{\, gen, \, \cH}$ \cite{agt}. In its original 
formulation, the AGT correspondence applies only to generic, that is non-minimal conformal field 
theories. From 4D instanton partition functions with $\epsilon_1 + \epsilon_2 = 0$, one obtains
conformal blocks of the conformal field theory of a Gaussian free field at the free fermion point.
From 4D instanton partition functions with $\epsilon_1 + \epsilon_2 \neq 0$, one obtains
conformal blocks of non-Gaussian, but also non-minimal conformal field theories.

\subsubsection{From generic conformal field theories to minimal models} One can choose the parameters 
that appear in $\cZ^{\, 4D}_{instanton}$, and restrict the states that are allowed as intermediate states, 
in such a way that one obtains restricted 4D instanton partition functions $\cZ^{\, 4D, \, min}_{instanton}$ 
that can be identified with conformal blocks in Virasoro $A$-series minimal models times Heisenberg factors 
\cite{alkalaev.belavin, bershtein.foda}. In particular, as we will show in the sequel, choosing 
$\epsilon_1 = - \sqrt{   p / \pp}$, and 
$\epsilon_2 =   \sqrt{ \pp /   p}$, 
where $p$ and $\pp$ are two co-prime positive integers, $0 < p < \pp$, one obtains conformal blocks in 
a minimal model parameterised by $p$ and $\pp$. 

\subsection{Plan of this work}
From the above chain of connections, it is expected that one can start from $\cZ^{\, ref\, top}$, 
choose the parameters and restrict the intermediate states to obtain 5D instanton partition functions 
$\cZ^{\, 5D, \, min}_{instanton}$, reduce to the corresponding $\cZ^{\, 4D, \, min}_{instanton}$, and 
compute minimal model conformal blocks from the latter. 
In this note, we work out the above chain of connections, which amounts to extending the result 
of \cite{alkalaev.belavin, bershtein.foda} by starting from topological strings and topological 
vertices rather than from 4D instanton partition functions. We glue four refined topological 
vertices to obtain the building block of the 5D instanton partition functions, and take the 4D 
limit of the latter to obtain the building block of the 4D instanton partition functions
$\cZ^{\, 4D}_{building.block}$. From that point on, we use the results of \cite{alkalaev.belavin, 
bershtein.foda} to obtain conformal blocks in minimal models.

Since $\cZ^{\, 4D}_{building.block}$, which can be regarded as the starting point of the results 
in \cite{alkalaev.belavin, bershtein.foda}, is constructed here from refined topological vertices, 
the construction in this note is, in this sense, more basic than that in \cite{alkalaev.belavin, 
bershtein.foda}. 

The Virasoro $A$-series minimal conformal blocks that we obtain can be computed using other methods, 
but we view this note as an accessible introduction to one more approach to the minimal conformal 
blocks that, hopefully, can be extended to minimal blocks beyond what can currently be computed, 
including the $W_N$ minimal blocks that do not satisfy the conditions of \cite{fateev.litvinov, wyllard}
\footnote{\,
Expressions for the 3-point functions of $W_N$ Toda theory were proposed, starting from topological 
strings, in \cite{mitev.01}, and checked in \cite{mitev.02}. These results should be useful in 
subsequent studies of $W_N$ minimal models.
}. 

\subsection{Outline of contents}
To simplify the presentation, we divide the content into short sections.
In {\bf \ref{section.02.refined.topological.vertex}}, we recall basic definitions related to Young 
diagrams and Schur functions, followed by
the definition of the refined topological vertex $\cC_{\, \lambda \, \mu \, \nu}[q, t]$ of Iqbal, 
Kozcaz and Vafa \cite{ikv}, where $\lambda$, $\mu$, and $\nu$ are Young diagrams, $q$ and $t$ are 
parameters.
In {\bf \ref{section.03.5d.web}}, we glue four copies of $\cC_{\, \lambda \, \mu \, \nu} [q, t]$ 
to build a $U(2)$ basic web partition function $\cW_{\, \bf V \, W \, \Delta} [q, t, R]$, where 
each of ${\bf V}$ and ${\bf W}$ is a pair of Young diagrams, and 
${\bf \Delta} = \{\Delta_1, \Delta_{\MM}, \Delta_2 \}$ are K{\"a}hler parameters.
In {\bf \ref{section.04.4d.web}}, we set $q= e^{\, R \et}$, $t= e^{\, - R \eo}$, where $\eo$ and 
$\et$ are Nekrasov's regularisation parameters, then take the limit $R \rightarrow 0$, to obtain 
$\cW_{\, \bf V \, W \, \Delta} [\epsilon_1, \epsilon_2, R \! \rightarrow \! 0]$.
In {\bf \ref{section.05.nekrasov.partition.functions}}, we recall the definition of the normalised 
contribution of the bifundamental hypermultiplet, 
$\cZ_{building.block}^{\, 4D}$ to  $\cZ^{\, 4D}_{instanton}$. 
In {\bf \ref{section.06.identification.numerators}}, we compare the numerators of 
$\cW^{\, norm}_{\bf V, W, \Delta}$ and $\cZ^{\, 4D}_{building.block}$, and identify the K{\"a}hler 
parameters with gauge theory parameters.
In {\bf \ref{section.07.identification.denominators}}, we compare the denominators of 
$\cW^{\, norm}_{\bf V, W, \Delta}$ and $\cZ^{\, 4D}_{building.block}$, and show that there is 
a normalisation such that the two denominators agree, without changing the results of the computations 
of the instanton partition functions.
In {\bf \ref{section.08.restricted.instanton.partition.functions}}, we recall the choice of parameters 
that allows us to use $\cZ^{\, 4D}_{building.block}$ to build 4D instanton partition functions that can 
be identified with Virasoro $A$-series minimal model conformal blocks times Heisenberg factors.
In {\bf \ref{section.09.from.gauge.theory.to.minimal.model}}, we identify the parameters of 
$\cW^{\, norm}_{\bf V \, W \, \Delta}$ and of $\cZ^{\, 4D}_{building.block}$. 
In {\bf \ref{section.10.burge.pairs}}, we outline a proof, following \cite{bershtein.foda}, of 
the statement that we need to impose Burge conditions on the partition pairs that appear in our 
constructions, to make the topological string partition functions free of non-physical singularities, 
once we choose the parameters to coincide with those of the minimal model.
Section {\bf \ref{section.11.comments}} includes comments and remarks. 
\subsection*{Abbreviations and notation} 
We focus on $U(2)$ quiver gauge theories, and simply say 
{\it \lq the instanton partition functions\rq\,}.
We assume that every conformal block, whether Liouville or minimal, 
includes a factor from a field theory 
of a free boson on a line, and omit {\it \lq times Heisenberg factors\rq}. 
The normalised contribution of the bifundamental hypermultiplet is understood as the 
building block of the instanton partition function, as all other contributions can be 
obtained from it. We simply say {\it \lq the bifundamental partition function\rq\,}.

\subsubsection*{Topological string-related notation}
$\cC_{\,     \lambda \, \mu \, \nu} [q, t]$ is the refined topological vertex.
$\cW_{\, \bf V \, W \, \Delta} [q, $ $t, R]$ is a basic web refined topological 
string partition function.
${\bf V}$ and ${\bf W}$ are two pairs of Young diagrams. 
${\bf \Delta}$ is a set of three K{\"a}hler parameters. 
$q$ and $t$ are deformation parameters. 
$R$ is the radius of a space-like circle
\footnote{\,
The 5D $U(2)$ quiver theories in this note live on D5-branes in time-like $x^0$, and 
space-like $x^1, x^2, x^3, x^5$ and $x^6$ dimensions. Following \cite{bao.01}, we take 
$x^5$ to be a circle of radius $R$, $\beta$ in \cite{bao.01}, such that setting 
$R \rightarrow 0$ is equivalent to taking the 4D limit. For a complete discussion, we 
refer to \cite{bao.01}.
}. 
$\cZ^{\, ref\, top}$ is the topological string partition function.

\subsubsection*{Gauge theory-related notation}
$\cZ^{\, 5D}_{instanton}$ is the 5D $U(2)$ quiver gauge theory instanton partition functions.
$\cZ^{\, 4D}_{instanton}$ is the 4D $U(2)$ quiver gauge theory instanton partition functions.
$\cZ_{building.block}^{\, 4D}$ is the normalised contribution of the bifundamental hypermultiplet.
The parameters $\eo$ and $\et$ are deformation parameters.

\subsubsection*{Conformal field theory-related notation}
$\cB^{\, gen, \,  \cH}$ are 2D Virasoro generic conformal blocks times Heisenberg factors.
$\cB^{\, gen, \,  min}$ are 2D Virasoro A-series minimal model conformal blocks times Heisenberg 
factors.
The parameters $p$ and $\pp$ are coprime positive integers that label a Virasoro minimal model. 
The parameters $r$ and $s$ are integers that satisfy $0 < r < p$, $0 < s < \pp$ and label Virasoro 
minimal model highest weight representations.

\section{The refined topological vertex}
\label{section.02.refined.topological.vertex}
{\it We recall basic definitions related to Young diagrams and Schur functions, followed by the 
definition of the refined topological vertex.}

\subsection{Partitions and Young diagrams}

\subsubsection{Partitions}
\label{partitions} A partition $\pi$ of a non-negative integer $|\pi|$
is a set of non-negative integers $\{\pi_{1},\pi_{2},$ $\cdots,\pi_{p}\}$,
where $p$ is the number of parts, $\pi_{i}\geq\pi_{i+1}$, and $\sum_{i=1}^{p} \pi_{i}= | \pi |$.

\subsubsection{Young diagrams}
\label{young.diagrams} 
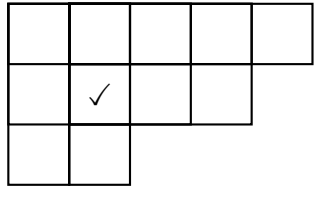
\begin{figure}
\begin{tikzpicture}[scale=.8]
\draw [thick] (0, 0) rectangle (1,1);
\draw [thick] (1, 0) rectangle (1,1);
\draw [thick] (2, 0) rectangle (1,1);
\draw [thick] (3, 0) rectangle (1,1);
\draw [thick] (4, 0) rectangle (1,1);
\draw [thick] (5, 0) rectangle (1,1);
\draw [thick] (0,-1) rectangle (1,1);
\draw [thick] (1,-1) rectangle (1,1);
\draw [thick] (2,-1) rectangle (1,1);
\draw [thick] (3,-1) rectangle (1,1);
\draw [thick] (4,-1) rectangle (1,1);
\draw [thick] (0,-2) rectangle (1,1);
\draw [thick] (1,-2) rectangle (1,1);
\draw [thick] (2,-2) rectangle (1,1);
\node at (1.5,-0.5) {$\checkmark$};
\end{tikzpicture}
\caption{
{\it
The Young diagram $Y$ of $5 + 4 + 2$.
The rows are numbered from top to bottom. 
The columns are numbered from left to right. 
The checkmarked cell has 
$A^{  } = 2$, 
$A^{+ } = \frac52$, 
$A^{++} = 3$, 
$L^{  } = 1$, {\it etc.}
}
}
\label{A.Young.diagram}
\end{figure}

A partition $\pi$ is represented as a Young diagram $Y$, as in 
Figure {\bf \ref{A.Young.diagram}}, which is a set of $p$ rows $\{y_1, y_2, \cdots, y_p \}$, such 
that row-$i$ has $y_i = \pi_{i}$ cells
\footnote{\, 
We use $y_i$ for $i$-th row as well as for the number of cells in that row.
}, 
$y_i \geq y_{i+1}$, and $\sum_i y_i = |Y| = |\pi|$. We use $Y^{\intercal}$ for the transpose of $Y$.

\subsubsection{Cells}

We use $\ws$ for a cell, or a square in the south-east quadrant
of the plane, and refer to the coordinates of $\ws$ as $\{\rrho,\ssigma\}$.
If $\ws$ is inside a Young diagram $Y$, then $\rrho$ is the
$Y$-row-number, counted from top to bottom, and $\ssigma$ is the
$Y$-column-number, counted from left to right, that $\ws$ lies
in. If $\ws$ is outside $Y$, we still regard $\rrho$ as a $Y$-row-number,
albeit the length of this row is zero, and we still regard $\ssigma$
as a $Y$-column-number, albeit the length of this column is zero.

\subsubsection{Arms and legs, half-extended and extended}
\label{arms.legs} Consider a cell $\ws$ that has coordinates $\{ \rrho, \ssigma \}$. 
We define the lengths of 
the           arm $A^{  }_{\ws, Y}$, 
half-extended arm $A^{+ }_{\ws, Y}$, 
     extended arm $A^{++}_{\ws, Y}$, 
the           leg $L^{  }_{\ws, Y}$,
half-extended leg $L^{+ }_{\ws, Y}$, 
     extended leg $L^{++}_{\ws, Y}$, of $\ws$  
with respect to the Young diagram $Y$, to be

\begin{eqnarray}
A^{  }_{\ws, Y} & = & y_{\rrho}               - \, \ssigma, \quad
A^{+ }_{\ws, Y}   =   A^{  }_{\ws, Y} + \frac12,            \quad 
A^{++}_{\ws, Y}   =   A^{  }_{\ws, Y} +      1,  
\\
L^{  }_{\ws, Y} & = & y_{\ssigma}^{\intercal} - \, \rrho,   \quad
L^{+ }_{\ws, Y}   =   L^{  }_{\ws, Y} + \frac12,            \quad        
L^{++}_{\ws, Y}   =   L^{  }_{\ws, Y} +      1,
\end{eqnarray}

\subsubsection{Remark} $A_{\ws,Y}$ and $L_{\ws,Y}$ can be negative when $\ws$ lies 
outside $Y$. 

\subsubsection{Partition pairs}
\begin{figure}
\begin{tikzpicture}[scale=.8] 
\draw [thick] (0, 0) rectangle (1, 1); 
\draw [thick] (1, 0) rectangle (2, 1);
\draw [thick] (2, 0) rectangle (3, 1);
\draw [thick] (3, 0) rectangle (4, 1);
\draw [thick] (4, 0) rectangle (5, 1);
\draw [thick] (0,-1) rectangle (1, 0);
\draw [thick] (1,-1) rectangle (2, 0);
\draw [thick] (2,-1) rectangle (3, 0);
\draw [thick] (3,-1) rectangle (4, 0);
\draw [thick] (0,-2) rectangle (1,-1);
\draw [thick] (1,-2) rectangle (2,-1);
%
\draw [thick] (7, 0) rectangle (8, 1); 
\draw [thick] (8, 0) rectangle (9, 1); 
\draw [thick] (9, 0) rectangle (10,1);   
\draw [thick] (10,0) rectangle (11,1);  
\draw [thick] (7,-1) rectangle (8, 0); 
\draw [thick] (7,-2) rectangle (8,-1);       
\node at (3.5,-0.5) {$\checkmark$};
\node at (10.5,-0.5) {$\checkmark$};
\end{tikzpicture}
\caption{
{\it 
A partition pair $\{\mu, \nu\}$. $\mu$ is on the left, $\nu$ is on the right. 
The checkmarked cell at coordinates $(2, 4)$, in the lower right quadrant of 
the plane, is in $\mu$, but not in $\nu$, and has
$A^{  }_{\mu} = 0$,
$A^{+ }_{\mu} = \frac12$,
$A^{++}_{\mu} = 1$,
$L^{  }_{\mu} = 0$,
$L^{+ }_{\mu} = \frac12$,
$L^{++}_{\mu} = 1$,
as well as
$A^{  }_{\nu} = -3$, {\it etc.} and
$L^{  }_{\nu} = -1$, {\it etc.} 
}
}
\label{A.partition.pair}
\end{figure}
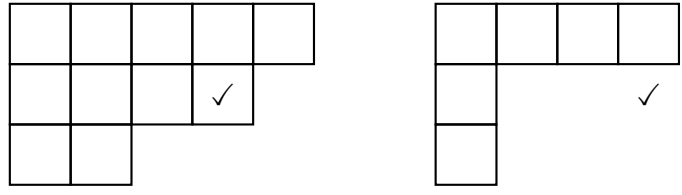

A partition pair ${\bf Y}$ is a set of two Young diagrams, $\{ Y^{\, 1}, Y^{\, 2} \}$, as in 
Figure {\bf \ref{A.partition.pair}}, where $|\bf Y| = |Y^{\, 1}| + |Y^{\, 2}|$ is the total 
number of cells in ${\bf Y}$.

\bigskip

\subsubsection{Sum of row-lengths and squares of row-lengths} Given 
a Young diagram $Y$, with rows of length $y_1 \geq y_2 \geq \cdots$, 
we define

\begin{equation}
| Y |                 = \sum_{i=1} y_i, 
\quad
\parallel Y \parallel = \sum_{i=1} y_i^{\, 2}
\end{equation}

\noindent where the sum is over all parts of $Y$
\footnote{\
In \cite{ikv}, Iqbal {\it et al.} use 
$\parallel Y \parallel^{\, 2} = \sum_{i=1} y_i^{\, 2}$. We define $\parallel Y \parallel = \sum_{i=1} y_i^{\, 2}$ 
to simplify the notation. This is consistent with other notation and should cause no confusion.
}.
Further, to simplify the equations, we define

\begin{equation}
| Y |^{\prime}                 = \frac12 \sum_{i=1} y_i,
\quad
\parallel Y \parallel^{\prime} = \frac12 \sum_{i=1} y_i^{\, 2}
\end{equation}

\subsubsection{Sequences} Given the sequence of row-lengths 
$  \lambda   = \{   \lambda_1,    \lambda_2, \cdots \}$, the sequence of half-integers
$  \rho      = \{ - \frac12,    - \frac32,   \cdots \}$, and two variables $x$ and $y$, 
we define the \lq exponentiated\rq\, sequences 

\begin{equation}
x^{-\lambda}= \{ x^{-\lambda_1}, x^{-\lambda_2}, \cdots \}, 
\quad
y^{\, -\rho}= \{ y^{\, \frac12}, y^{\, \frac32}, \cdots \},
\quad
\textit{and}
\quad
x^{\, -\lambda} y^{\, -\rho} = 
\{
x^{\, -\lambda_1} \, y^{\, \frac12}, 
x^{\, -\lambda_2} \, y^{\, \frac32}, 
\cdots \}, 
\end{equation}

\subsubsection{A function of the arm-lengths and the leg-lengths} Given a Young diagram 
$\lambda$, we define the function

\begin{equation}
\label{z.factor}
Z_{\lambda} [q, t]
= \prod_{\ws \in \lambda} 
\frac{1}{\ll 1 - \ll \frac{q}{t} \rr^{\, \frac12} \, 
q^{\, A_{\ws, \lambda}^{+ }} \, 
t^{\, L_{\ws, \lambda}^{+ }} 
\rr}
= \prod_{\ws \in \lambda} \frac{1}{\ll 1- \, q^{\, A_{\ws, \lambda}^{++}}\, 
                                             t^{\, L_{\ws, \lambda}^{  }} \rr}
\end{equation}

\subsubsection{Remark} 
The expression in the middle of Equation {\bf \ref{z.factor}} corresponds to 
splitting the cell $\ws$ at the corner of a hook in the partition $\lambda$ into two halves, then
attaching one       half to the arm of that hook to form {\it a half-extended arm} of length 
$A^{+ }_{\ws, \lambda} = A^{  }_{\ws, \lambda} + \frac12$, and 
attaching the other half to the leg of that hook to form {\it a half-extended leg} of length 
$L^{+ }_{\ws, \lambda} = L^{  }_{\ws, \lambda} + \frac12$. 
The expression on the right corresponds to 
attaching the cell $\ws$ at the corner of a hook to the arm of that hook to form 
{\it an extended arm} of length 
$A^{++}_{\ws, \lambda} = A^{  }_{\ws, \lambda} + 1$. The length of the leg of that hook remains 
$L^{  }_{\ws, \lambda}$. 

\subsection{Schur and skew-Schur functions}
Given an $n$-row Young diagram $\lambda$, with parts $\lambda_1 \geq \lambda_2 \geq \cdots$, 
and a set of $n$ variables $\{x_1, x_2, \cdots, x_n \}$, the Schur function $s_{\lambda} [{\bf x}]$
is defined as
\begin{equation}
s_{\lambda}[{\bf x}] =
\frac{
det \ll x_i^{\, \lambda_j+n-j} \rr_{1 \leq i, j \leq n}
}
{
\prod_{1 \leq i < j \leq n} \ll x_i - x_j \rr
}
\end{equation}
The skew-Schur function $s_{\lambda/\mu}[{\bf x}]$ is defined as
\begin{equation}
s_{\lambda/\mu}[{\bf x}]=\sum_{\nu}c_{\mu \nu}^{\, \lambda}\ 
s_{\lambda}[{\bf x}]
\end{equation}
\noindent where $c_{\mu \nu}^{\, \lambda}$ are Littlewood-Richardson coefficients defined by
\begin{equation}
s_{\mu}\ 
s_{\nu}=\sum_{\lambda}\ c_{\mu \nu}^{\, \lambda}\ 
s_{\lambda}
\end{equation}

\subsection{Topological vertices}
Our story starts from A-model closed topological string theory on non-compact Calabi-Yau 
threefolds.
We cannot afford to review this vast subject and refer the reader to excellent available 
introduction, including \cite{marino.2005, marino.book}.

\subsubsection{The topological vertex of Aganagic {\it et al}} 
\begin{figure}
\begin{tikzpicture}
\draw [ultra thick]  (0.0,2.0)--(2.0,2.0);
\draw [ultra thick]  (2.0,2.0)--(2.0,4.0);
\draw [ultra thick]  (2.0,2.0)--(3.5,0.5);
\node [left]  at (2.0,3.0) {$\lambda$};
\node [left]  at (2.8,1.0) {$\mu$};
\node [above] at (1.0,2.0) {$\nu$};
\end{tikzpicture}
\caption{
{\it 
The topological vertex $C_{\lambda \, \mu \, \nu} [q, t]$ is trivalent and depends 
on two parameters $q$ and $t$. 
The segments are labelled by three partitions $\lambda, \, \mu$ and $\nu$, such that 
$\lambda$ is assigned to the vertical segment, $\mu$ is assigned to the segment that 
follows in a clockwise direction, and $\nu$ to the segment that follows. 
The \lq preferred leg\rq\, is labelled by $\nu$. 
} 
}
\label{A.topological.vertex}
\end{figure}
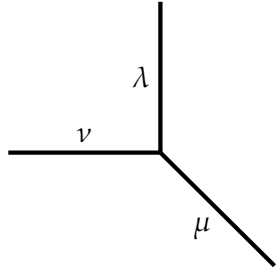

In \cite{topological.vertex}, Aganagic, Klemm, Marino and Vafa introduced a systematic procedure 
to calculate A-model topological string partition functions on resolved conifolds. The main 
ingredient of this procedure is the topological vertex, which has a combinatorial representation 
in terms of plane partitions with three boundaries specified by three Young diagrams, and can be 
schematically represented as a trivalent vertex, with bonds labelled by Young diagrams, as in 
Figure {\bf \ref{A.topological.vertex}}. For details, we refer to \cite{topological.vertex}.

\subsubsection{The refined topological vertex}
In \cite{ikv}, Iqbal, Kozcaz and Vafa defined the refined topological vertex, up to simple 
re-arrangements
\footnote{\,
We use 
$\cC_{\,     \lambda \, \mu \, \nu} [q, t]$, and
$Z_Y [q, t]$, while Iqbal {\it et al.} use
$\cC_{\, \lambda \, \mu \, \nu} [t, q]$, and
$\tilde{Z}_{\mu} [t, q]$.
}, as 

\begin{multline}
\label{refined.topological.vertex}
C_{\lambda \, \mu \, \nu} [q, t] =
\\
q^{\, \ll 
                | \lambda     |^{\prime} 
      -         | \mu         |^{\prime} 
      + \parallel \mu \parallel^{\prime} 
      + \parallel \nu \parallel^{\prime} 
    \rr} \, 
t^{\, \ll -     | {\lambda}^{         }         |^{\prime} 
      +         | {\mu    }^{         }         |^{\prime} 
      - \parallel {\mu    }^{\intercal} \parallel^{\prime}
\rr} \, 
Z_{\nu}[q, t]
\ll
\sum_{\eta} 
\ll 
\frac{q}{t} 
\rr^{\,| \eta |^{\prime}}
\ s_{ {\lambda}^{\intercal} / \eta}[ q^{\,- \nu } \ t^{\,- {\rho}^{         }}]
\ s_{ {\mu    }^{         } / \eta}[ q^{\,- \rho} \ t^{\,- {\nu }^{\intercal}}]
\rr
\end{multline}

\noindent Note that the refined vertex is not manifestly cyclically-symmetric, in the sense
that the partitions that label the external legs do not appear on equal footing. In particular, 
the partition $\nu$ is distinguished from the other two. The external leg labelled by $\nu$ is
referred to as {\it \lq preferred\rq\,}. The original, unrefined topological vertex of Aganagic 
{\it et al.} is recovered by setting $q=t$.

\subsubsection{On framing vectors and framing factors} 
In defining the refined topological vertex, one labels each of the three boundaries of a vertex 
by {\it a framing vector} that indicates a possible twisting of the boundary. On gluing two 
vertices along a common boundary, there is in general {\it a framing factor} that accounts for 
a possible mismatch in the orientations of the relevant framing vectors. In this note, we glue 
vertices such that we do not require framing factors.  

\section{A 5D $U(2)$ basic web partition function}
\label{section.03.5d.web}
{\it We glue four copies of the refined topological vertex to obtain a 5D basic web that can be 
used as a building block of $U(2)$ topological string partition functions.}

\subsection{Geometric engineering} Following \cite{geometric.engineering.01, geometric.engineering.02}, 
we write the normalised $U(2)$ basic web partition function as

\begin{equation}
\label{W.norm}
\cW_{\, \bf V \, W \, \Delta}^{\,  norm} [q, t, R] = 
\frac{
\cW_{\, \bf V \, W \, \Delta}^{        } [q, t, R]
}
{
\cW_{\, \bf \emptyset\, \emptyset\, \Delta}[q, t, R]
}
\end{equation}

\noindent where $\emptyset$ is the trivial, or empty partition with no cells. The numerator is

\begin{multline}
\label{the.numerator}
\cW_{\, \bf V \, W \, \Delta}[q, t, R]
=
\sum_{\xi_1, \, \xi_{\MM}, \, \xi_2}
\ll -Q_1     \rr^{\, |\xi_1     |} \,
\ll -Q_{\MM} \rr^{\, |\xi_{\MM} |} \,
\ll -Q_2     \rr^{\, |\xi_2  |} 
\\
\times
C_{\emptyset                  \, \xi_1             \, V^{\, 1}             } [q, t] \ 
C_{\xi_{\MM}               \, \xi_1^{\intercal} \, W^{\, 1 \, \intercal}} [t, q] \ 
C_{\xi_{\MM}^{\intercal}   \, \xi_2             \, V^{\, 2}             } [q, t] \ 
C_{\emptyset                  \, \xi_2^{\intercal} \, W^{\, 2 \, \intercal}} [t, q]
\end{multline}

\noindent where we use

\begin{equation}
\boxed{
Q_i = e^{\, - R \Delta_i}, \quad i = 1, {\MM}, 2
}
\end{equation}

\noindent and the denominator 
$\cW_{\, \bf \emptyset \, \emptyset \, \Delta}[q, t, R]$ 
is identical to the numerator 
$\cW_{\, \bf V \, W \, \Delta}[q, t, R]$
but with all external partition pairs empty.

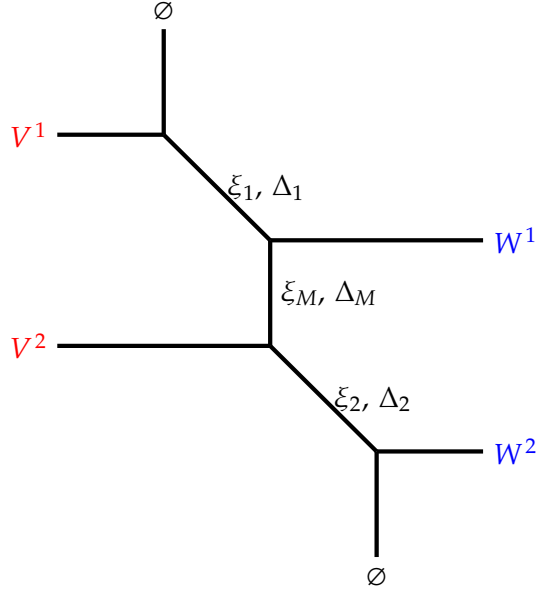
\begin{figure}
\begin{tikzpicture}[scale=.7]
\draw [ultra thick] (0,4)--(2,4);
\draw [ultra thick] (2,6)--(2,4);
\draw [ultra thick] (2,4)--(4,2);
\node [left, red] at (0,4) {$V^{\, 1}$};
\node [right] at (3,3) {$\xi_1, \, \Delta_1$};
\node [above] at (2,6) {$\emptyset$};
\draw [ultra thick] (4,2)--(8,2);
\draw [ultra thick] (4,0)--(4,2);
\node [right, blue] at (8,2) {$W^{\, 1}$};
\node [right] at (4,1) {$\xi_{\MM}, \, \Delta_{\MM}$};
\draw [ultra thick] (0,0)--(4,0);
\node [left, red] at (0,0) {$V^{\, 2}$};
\draw [ultra thick] (4,0)--(6,-2);
\node [right] at (5,-1) {$\xi_2, \, \Delta_2$};
\draw [ultra thick] (6,-2)--(6,-4);
\draw [ultra thick] (6,-2)--(8,-2);
\node [right,blue] at (8,-2) {$W^{\, 2}$};
\node [below] at (6,-4) {$\emptyset$};
\end{tikzpicture}
\caption{
{\it 
The $U(2)$ basic web diagram. Each external line is labeled by a partition. 
Each internal line is labeled by a partition and a K{\"a}hler parameter. This basic 
web can be glued to form topological partition functions. The preferred legs are all
external, and labelled by the partitions $V^1, V^2, W^1$ and $W^2$.
}
}
\label{basic.web}
\end{figure}

As shown in Figure {\bf \ref{basic.web}}, the basic web has two external horizontal legs 
coming in from the left, two external horizontal legs going out to the right, and a pair 
of vertical legs, one going up and one down.
The horizontal external legs on the left  are assigned partitions $\{V^{\, 1}, V^{\, 2}\}$, 
the horizontal external legs on the right are assigned partitions $\{W^{\, 1}, W^{\, 2}\}$. 
The internal lines are assigned parameters $  Q_1$, $  Q_{\MM}$ and $  Q_2$, and partitions 
                                           $\xi_1$, $\xi_{\MM}$ and $\xi_2$,
from top to bottom. 
The vertical external legs are assigned empty partitions. 

Each trivalent vertex corresponds to a refined topological vertex $\cC_{\lambda \, \mu \, \nu}$. 
Our convention is such that each vertex has one vertical leg, we associate $\lambda$ to that 
vertical leg, regardless of whether it is internal or external, pointing upwards or downwards, 
then $\mu$ and $\nu$ to the remaining two legs, encountered sequentially as we start from the 
vertical leg and move around the vertex clockwise. 
Using Equation {\bf \ref{refined.topological.vertex}}, we re-write the numerator in Equation 
{\bf \ref{the.numerator}} as

\begin{multline}
\cW_{\, \bf V \, W \, \Delta}^{        } [q, t, R]
\\
q^{\, \ll \parallel    V^{\, 1}              \parallel^{\prime} + \parallel   V^{\, 2}              \parallel^{\prime} \rr}
t^{\, \ll \parallel    W^{\, 1 \, \intercal} \parallel^{\prime} + \parallel   W^{\, 2 \, \intercal} \parallel^{\prime} \rr}
Z_{  V^{\, 1}             } [q, t]\
Z_{  W^{\, 1 \, \intercal}} [t, q]\
Z_{  V^{\, 2}             } [q, t]\
Z_{  W^{\, 2 \, \intercal}} [t, q]
\\
\times \,
\sum_{\, \xi_1, \, \xi_{\MM}, \, \xi_2, \, \eta_1, \, \eta_2}
\ll - Q_1     \rr^{\, |\xi_1     |} 
\ll - Q_{\MM} \rr^{\, |\xi_{\MM} |}
\ll - Q_2     \rr^{\, |\xi_2     |}
\ll \frac{q}{t} \rr^{\, \ll | \eta_2 |^{\prime} - |\eta_1 |^{\prime} \rr}
\\
\times
s_{\xi_1                         } \, [q^{\, -\rho} \, t^{\, -V^{\, 1 \, \intercal}}]     \ 
s_{\xi_1^{\intercal}     / \eta_1} \, [q^{\, - W^{\, 1}} \, t^{\, -\rho                 }]\
s_{\xi_{\MM}^{\intercal} / \eta_1} \, [q^{\, -\rho     } \, t^{\, -W^{\, 1 \, \intercal}}]\ 
\\
\times
s_{\xi_{\MM}             / \eta_2} \, [q^{\, - V^{\, 2}} \, t^{\, -\rho           }]      \
s_{\xi_2                 / \eta_2} \, [q^{\, -\rho}      \, t^{\, -V^{\, 2 \, \intercal}}]\ 
s_{\xi_2^{\intercal}             } \, [q^{\, - W^{\, 2}} \, t^{\, -\rho           }]      \ 
\label{web.02}
\end{multline}

\noindent where we used the fact that for an empty partition $\emptyset$, the skew 
partition $\emptyset / \eta$ exists only for $\eta = \emptyset$, the sum over $\eta$ 
trivialises, and the skew Schur function $s_{\emptyset / \eta} = s_{\emptyset} = 1$.

\subsubsection{Two skew Schur function identities}
To evaluate the sums in Equation {\bf \ref{web.02}} for $w^{\, num}$, we need the two identities
\footnote{\,
Exercise {\bf 26}, page 93 of \cite{macdonald.book}.}

\begin{eqnarray}
\sum_{\lambda} \, s_{\lambda / \eta_1}[x] \, s_{\lambda / \eta_2}[y] 
& = & 
\prod_{i, j} \ll 1-x_i y_j \rr^{-1} 
\sum_{\tau}  \, s_{\eta_1 / \tau}[y] \, s_{\eta_2 / \tau} [x]
\label{schur.01}
\\
\sum_{\lambda} \, s_{\lambda / \eta_1}[x] \, s_{\lambda^{\intercal} / \eta_2}[y] & = & 
\prod_{i, j} \ll 1+x_i y_j \rr      
\sum_{\tau}  \, s_{\eta_1^{\intercal} / \tau^{\intercal}}[y] \, s_{\eta_2^{\intercal}/ \tau} [x]
\label{schur.02}
\end{eqnarray}

\noindent as well as the property that

\begin{equation}
Q^{\, |\lambda| - |\eta| } \, s_{\lambda/\eta}[x]= s_{\lambda / \eta} \, [Q x]
\end{equation}

\noindent which follows from the definition of the skew Schur function. 

\subsubsection{The basic web in product form}
We evaluate the sums over the right hand side of Equation {\bf \ref{web.02}}, using 
$\rho_i = - i + \frac12$.

\subsubsection{The sums over $\xi_1$ and $\xi_2$}

\begin{multline}
\sum_{\xi_1} \ll - Q_1 \rr^{\,|\xi_1|}  \, 
s_{\xi_1                     } [q^{\, -\rho} \, t^{\, -V^{\, 1 \, \intercal}}]\, 
s_{\xi_1^{\intercal} / \eta_1} [q^{\, - W^{\, 1}} \, t^{\, -\rho           }]
\\
= \prod_{i, j = 1}^{\infty}
\ll
                      1 - Q_1 \, q^{\, -W_i^{\, 1} + j - \frac12} \, t^{\, -V_j^{\, 1 \, \intercal} + i - \frac12}
\rr \, 
s_{\eta_1^{\intercal}} [- Q_1 \, q^{\, -\rho}                \, t^{\, -V^{\, 1 \, \intercal}}],
\label{4.3}
\end{multline}

\begin{multline}
\sum_{\xi_2}  \,
\ll -Q_2 \rr^{\, |\xi_2|} \, 
s_{\xi_2 / \eta_{2}} [ q^{\, -\rho} \, t^{\, - V^{\, 2 \, \intercal}}] \, 
s_{\xi_2^{\intercal}}[ q^{\, - W^{\, 2}} \, t^{\, - \rho           }]
\\
=\prod_{i,j=1}^{\infty} \,
\ll
1-Q_2 \, q^{\, -W_i^{\, 2             }+j-\frac12} 
      \, t^{\, -V_j^{\, 2 \, \intercal}+i-\frac12}
\rr \ 
s_{\eta_2^{\intercal} }[-Q_2 q^{\, -W^{\, 2}} \, t^{\, -\rho} ]
\label{4.4}
\end{multline}

\subsubsection{The sum over $\xi_{\MM}$} We re-write this in terms of a sum over 
a new set of partition $\tau$,

\begin{multline}
\sum_{\xi_{\MM} } \, 
\ll -Q_{\MM} \rr^{\, | \xi_{\MM}  | - | \eta_1 | } \,
s_{\xi_{\MM}^{\intercal} / \eta_1} [ \, q^{\, -\rho} \, t^{\, -W^{\, 1, \, \intercal}} ] \, 
s_{\xi_{\MM}             / \eta_2} [ \, q^{\, -V^{\, 2}}  \, t^{\, -\rho}            ]
\\
=\prod_{i, j}
\ll
1 - Q_{\MM} \, 
q^{\, -V_i^{\, 2             } + j - \frac12} \,
t^{\, -W_j^{\, 1 \, \intercal} + i - \frac12}
\rr
\sum_{\tau}
\ 
s_{\eta_1^{\intercal}/ \tau^{\intercal}}[             q^{\, -V^{\, 2} } \, t^{\, -\rho}            ] \, 
s_{\eta_2^{\intercal}/ \tau            }[-Q_{\MM}  \, q^{\, -\rho} \, t^{\, -W^{\, 1 \, \intercal}} ]
\label{4.5}
\end{multline}

\subsubsection{The sums over $\eta_1$ and $\eta_2$} 

\begin{multline}
\sum_{\eta_1}
\ll \frac{q}{t} \rr^{\, - |\eta_1|^{\prime}} \, 
\ll - Q_{\MM}\rr^{  |\eta_1|         } \,
s_{\eta_1^{\intercal}}                     [      -Q_1 \, q^{\, -\rho} \, t^{\, -V^{\, 1 \, \intercal}}]\ 
s_{\eta_1^{\intercal} / \tau^{\intercal}} \, [            q^{\, - V^{\, 2}} \, t^{\, -\rho           }]
\\
=\prod_{i,j}
\ll
1-Q_1 \, Q_{\MM}\, q^{\, -V_i^{\, 2} + j - 1} \, t^{\, -V_j^{\, 1 \, \intercal}+i} 
\rr^{-1}
\ 
s_{\tau^\intercal}[Q_1 \, Q_{\MM} \, q^{\, j - \frac12} \, t^{\, -V_j^{\, 1 \, \intercal}}]
\label{4.6}
\end{multline}

\begin{multline}
\sum_{\eta_2} 
\ll \frac{q}{t} \rr^{\,  |\eta_2|^{\prime}} \,
s_{\eta_2^\intercal       }[-Q_2     \, q^{\, - W^{\, 2}} \, t^{\, -\rho                 } ] \, 
s_{\eta_2^\intercal / \tau}[-Q_{\MM} \, q^{\, -\rho}      \, t^{\, -W^{\, 1 \, \intercal}} ]
\\
=\prod_{i,j}
\ll      1 - Q_{\MM} \, Q_2 \,  q^{\, -W_i^{\, 2} + j} \, t^{\, -W_j^{\, 1 \, \intercal} + i - 1} \rr^{-1} \, 
s_{\tau}[  - Q_2     \,         q^{\, -W_i^{\, 2}    } \, t^{\, i - \frac12} ]
\label{4.7}
\end{multline}

\subsubsection{The sum over $\tau$} We finally evaluate the sum over the partitions 
that were introduced in an intermediate step above, 

\begin{equation}
\sum_{\tau}\ 
s_{\tau}[-Q_2 \,                   q^{\, -W_i^{\, 2}} \, t^{\, i - \frac12         }] \, 
s_{\tau^{\intercal}}[Q_1 \, Q_{\MM} \, q^{\, j - \frac12} \, t^{\, - V_j^{\, 1 \, \intercal} }]
=\prod_{i, j}
\ll 1 - Q_1 \, Q_{\MM} \, Q_2 \, q^{\, - W_i^{\, 2} + j - \frac12} t^{\, -V_j^{\, 1 \, \intercal} + i - \frac12} \rr
\label{4.8}
\end{equation}

\noindent to obtain 

\begin{multline}
\label{w.num}
\cW_{\, \bf V \, W \, \Delta}^{        } [q, t, R]
= 
\\
q^{\, \ll \parallel    V^{\, 1             } \parallel^{\prime} + \parallel   V^{\, 2             } \parallel^{\prime} \rr}
t^{\, \ll \parallel    W^{\, 1 \, \intercal} \parallel^{\prime} + \parallel   W^{\, 2 \, \intercal} \parallel^{\prime} \rr} \, 
Z_{  V^{\, 1}             }[q, t]\
Z_{  W^{\, 1 \, \intercal}}[t, q]\
Z_{  V^{\, 2}             }[q, t]\
Z_{  W^{\, 2 \, \intercal}}[t, q]
\\
\times \,
\prod_{i, j = 1}^{\infty} \, \ll 1  - Q_1         \, 
q^{\, -W_i^{\, 1} + j - \frac12}                  \, 
t^{\, -V_j^{\, 1 \, \intercal} + i - \frac12} \rr \, 
\prod_{i, j = 1}^{\infty} \, \ll 1 - Q_2          \, 
q^{\, -W_i^{\, 2}             + j - \frac12}      \, 
t^{\, -V_j^{\, 2 \, \intercal} + i - \frac12} \rr \,
\\
\times \,
\prod_{i, j = 1}^{\infty} \, \ll 1 -        Q_{\MM}        \,               
q^{\, -V_i^{\, 2} + j - \frac12} \, t^{\, - W_j^{\, 1 \, \intercal} + i - \frac12} \rr \,
\
\prod_{i, j = 1}^{\infty} \, \ll 1 - Q_1 \, Q_{\MM} \, Q_2 \, 
q^{\, -W_i^{\, 2} + j - \frac12} \, t^{\, - V_j^{\, 1 \, \intercal} + i - \frac12} \rr
\\
\times \,
\prod_{i, j = 1}^{\infty} \, \ll 1 - Q_1 \, Q_{\MM} \,        
q^{\, - V_i^{\, 2}             + j - 1} \, 
t^{\, - V_j^{\, 1 \, \intercal} + i    } \rr^{-1} 
\
\prod_{i, j = 1}^{\infty} \, \ll 1 -        Q_{\MM} \, Q_2 \,        
q^{\, - W_i^{\, 2}             + j    } \, 
t^{\, - W_j^{\, 1 \, \intercal} + i - 1} \rr^{-1} 
\end{multline}

\noindent Note that regarding the right hand side of Equation {\bf \ref{w.num}} as a rational function, 
the initial four products are in the numerator, while the latter two are in the denominator. 

\subsubsection{Normalised products}
Given two partitions, $V$ with parts $v_i$, $i = 1, 2, \cdots$, and 
                      $W$ with parts $w_i$, $i = 1, 2, \cdots$, 
and two sequences of integers $\alpha_k$ and $\beta_k$, $k = 1, 2, \cdots$, we define 

\begin{equation}
\label{normalised.product}
{\prod_{i, j=1}^{\infty}}^{^\prime} 
\ll 1-Q \, q^{\, - v_i + \alpha_j} \, t^{\, - w_j + \beta_i} \rr
=
\prod_{i, j=1}^{\infty}
\ll
\frac{
1-Q \, q^{\, - v_i + \alpha_j} \, t^{\, - w_j + \beta_i}
}{
1-Q \, q^{\,         \alpha_i} \, t^{\,         \beta_i}
}
\rr
\end{equation}

\noindent In this notation, the expression for 
$\cW_{\, \bf V \, W \, \Delta}^{\, norm} [q, t, R]$ 
in Equation {\bf \ref{W.norm}}, is identical to that for 
$\cW_{\, \bf V \, W \, \Delta}[q, t, R]$, 
in Equation {\bf \ref{web.02}}, up to replacing each product $\prod_{i, j}$ 
by a normalised product $\prod_{i, j}^{^\prime}$.

\subsubsection{From infinite to finite products}

\noindent Using Equation {\bf \ref{normalised.product}}, we have the following identities 
\cite{nakajima.yoshioka, awata.kanno.02}
\footnote{\
For an excellent reference and compendium of relevant combinatorial identities, including 
proofs, see \cite{awata.kanno.02}. Equation {\bf \ref{nakajima.identity}} in this note 
follows from Equations {\bf 2.8--2.11} in \cite{awata.kanno.02}.
}. Firstly,

\begin{eqnarray}
\label{nakajima.identity}
{\prod_{i, j=1}^{\infty}}^{^\prime} 
\ll 1-Q \, q^{\, -w_i + j - 1} \, t^{\, -v_j^{\intercal} + i} \rr
& = & 
\prod_{\ws \in V}
\ll
1-Q\, q^{\, -A_{\ws, V}^{++}} \, t^{\, -L_{\ws, W}^{  }}
\rr
\prod_{\bs \in W}
\ll
1-Q\, q^{\,  A_{\bs, W}^{  }} \, t^{\,  L_{\bs, V}^{++}}
\rr
\nonumber
\\
& = & 
\prod_{\bs \in W}
\ll
1-Q\, q^{\, -A_{\bs, V}^{++}} \, t^{\, -L_{\bs, W}^{  }}
\rr
\prod_{\ws \in V}
\ll
1-Q\, q^{\,  A_{\ws, W}^{  }} \, t^{\,  L_{\ws, V}^{++}}
\rr
\end{eqnarray}

\noindent Note that while the product on the left of Equation {\bf \ref{nakajima.identity}} is 
normalised in the sense of Equation {\bf \ref{normalised.product}}, the remaining products are 
not.  

\subsubsection{The 5D basic web in product form}

Using the identities {\bf \ref{nakajima.identity}} in Equation {\bf \ref{web.02}} for the numerator, 
we write the normalised $U(2)$ basic web partition function in Equation {\bf \ref{W.norm}} as

\begin{equation}
\label{W.5D.norm}
\cW^{\, norm}_{\, \bf V \, W \, \Delta}[q, t, R]
=
q^{\, \ll \parallel    V^{\, 1}              \parallel^{\prime} + \parallel   V^{\, 2}              \parallel^{\prime} \rr}
t^{\, \ll \parallel    W^{\, 1 \, \intercal} \parallel^{\prime} + \parallel   W^{\, 2 \, \intercal} \parallel^{\prime} \rr}
\
\frac{
w^{\, num} [{\bf V, \, W, \, \Delta}, q, t, R]
}{
w_{\, den} [{\bf V, \, W, \, \Delta}, q, t, R]
}
\end{equation}

\noindent where, using $Q_3 = Q_1 Q_{\MM} Q_2$, $w^{num}$ is 

\begin{eqnarray}
\label{w.numerator.5D}
w^{\, num} [{\bf V, \, W, \, \Delta}, q, t, R] 
& = & 
\prod_{\ws \in V^{\, 1}} \ll 1-Q_1    \, q^{\, -A_{\ws, W^{\, 1}}^+} \, t^{\, -L_{\ws, V^{\, 1}}^+} \rr
\prod_{\bs \in W^{\, 1}} \ll 1-Q_1    \, q^{\,  A_{\bs, V^{\, 1}}^+} \, t^{\,  L_{\bs, W^{\, 1}}^+} \rr
\nonumber
\\
& \times &
\prod_{\ws \in V^{\, 2}} \ll 1-Q_2    \, q^{\, -A_{\ws, W^{\, 2}}^+} \, t^{\, -L_{\ws, V^{\, 2}}^+} \rr 
\prod_{\bs \in W^{\, 2}} \ll 1-Q_2    \, q^{\,  A_{\bs, V^{\, 2}}^+} \, t^{\,  L_{\bs, W^{\, 2}}^+} \rr
\nonumber
\\
& \times &
\prod_{\bs \in W^{\, 1}} \ll 1-Q_{\MM} \, q^{\, -A_{\bs, V^{\, 2}}^+} \, t^{\, -L_{\bs, W^{\, 1}}^+} \rr
\prod_{\ws \in V^{\, 2}} \ll 1-Q_{\MM} \, q^{\,  A_{\ws, W^{\, 1}}^+} \, t^{\,  L_{\ws, V^{\, 2}}^+} \rr
\nonumber
\\
& \times &
\prod_{\ws \in V^{\, 1}} \ll 1- Q_3    \, q^{\, -A_{\ws, W^{\, 2}}^+} \, t^{\, -L_{\ws, V^{\, 1}}^+} \rr 
\prod_{\bs \in W^{\, 2}} \ll 1- Q_3    \, q^{\,  A_{\bs, V^{\, 1}}^+} \, t^{\,  L_{\bs, W^{\, 2}}^+} \rr, 
\end{eqnarray}

\noindent where we have used the second equality in Equation {\bf \ref{nakajima.identity}} 
to put the products in the above uniform form. The denominator $w_{den}$ is 

\begin{eqnarray}
\label{w.denominator.5D}
w_{\, den} [{\bf V, \, W, \, \Delta}, q, t, R] 
& = &
\prod_{\ws \in V^{\, 1}} \ll 1-                 q^{\, A_{\ws,   V^{\, 1}}^{++}} \, t^{\, L_{\ws, V^{\, 1}}^{  }} \rr
\prod_{\bs \in W^{\, 1}} \ll 1-                 q^{\, A_{\bs,   W^{\, 1}}     } \, t^{\, L_{\bs, W^{\, 1}}^{++}} \rr
\nonumber
\\
& \times &
\prod_{\ws \in V^{\, 2}} \ll 1-                 q^{\, A_{\ws,   V^{\, 2}}^{++}} \, t^{\, L_{\ws, V^{\, 2}}^{  }} \rr
\prod_{\bs \in W^{\, 2}} \ll 1-                 q^{\, A_{\bs,   W^{\, 2}}     } \, t^{\, L_{\bs, W^{\, 2}}^{++}} \rr
\nonumber
\\
& \times &
\prod_{\ws \in V^{\, 1}} \ll 1-Q_1 \, Q_{\MM} \, q^{\, -A_{\ws, V^{\, 2}}^{++}} \, t^{\, -L_{\ws, V^{\, 1}}     } \rr
\prod_{\ws \in V^{\, 2}} \ll 1-Q_1 \, Q_{\MM} \, q^{\,  A_{\ws, V^{\, 1}}     } \, t^{\,  L_{\ws, V^{\, 2}}^{++}} \rr 
\nonumber
\\
& \times &
\prod_{\bs \in W^{\, 1}} \ll 1-Q_{\MM} \, Q_2 \, q^{\, -A_{\bs, W^{\, 2}}     } \, t^{\, -L_{\bs, W^{\, 1}}^{++}} \rr
\prod_{\bs \in W^{\, 2}} \ll 1-Q_{\MM} \, Q_2 \, q^{\,  A_{\bs, W^{\, 1}}^{++}} \, t^{\,  L_{\bs, W^{\, 2}}     } \rr
\end{eqnarray}

\noindent where the first four products on the right hand side of Equation {\bf \ref{w.denominator.5D}} 
are due to the product 
$ Z_{  V^{\, 1}                  }[t, q]\,
  Z_{  W^{\, 1 \, \intercal}     }[q, t]\, 
  Z_{  V^{\, 2}                  }[t, q]\, 
  Z_{  W^{\, 2 \, \intercal}     }[q, t]$ in Equation {\bf \ref{web.02}}.
$w_{den}$ is equal to the denominator $\cW_{\, \bf \emptyset \, \emptyset \, \Delta} [q, t, R]$ on the right hand 
side of Equation {\bf \ref{W.norm}}.

\subsubsection{Remark} One can glue copies of the basic web partition function 
$\cW_{\, \bf V \, W \, \Delta}[q, t, R]$ in several ways. In this note, we 
restrict our attention to gluing linearly or cyclically, to form linear or 
cyclic $U(2)$ quiver gauge theories, as described in paragraphs 
{\bf \ref{linear.conformal.blocks}} and {\bf \ref{cyclic.conformal.blocks}}.
We do not, for example, glue basic webs to form a Hirzbruch surface.

\section{A 4D $U(2)$ basic web partition function}
\label{section.04.4d.web}
{\it 
We take the $R \! \rightarrow \! 0$ limit of the 5D basic web partition function to obtain 
its 4D analogue. 
}

\subsection{Two parameters} We take the relationship between the parameters $q$ and $t$ of 
the refined vertex and the parameters $\epsilon_1$ and $\epsilon_2$ of the instanton 
partition function to be,
\begin{equation}
\boxed{
q=e^{\,   R \et},
\quad
t=e^{\, - R \eo}
}
\end{equation} 

\noindent where $R$, the radius of a space-like circle, plays the role of a deformation parameter. 
We write $\cW_{\, \bf V \, W \, \Delta}^{\, norm} [\epsilon_1, \epsilon_2, R]$, then take the limit 
$R \! \rightarrow \! 0$. The prefactor on the left hand side of Equation {\bf \ref{W.5D.norm}} tends 
to 1 in the limit $R \rightarrow 0$, and we obtain

\begin{equation}
\cW_{\, \bf V \, W \, \Delta}^{\, norm} [\epsilon_1, \epsilon_2, R \rightarrow 0] 
=
\frac{
w^{\, num} [{\bf V, \, W, \, \Delta}, \epsilon_1, \epsilon_2, R \rightarrow 0] 
}{
w_{\, den} [{\bf V, \, W, \, \Delta}, \epsilon_1, \epsilon_2, R \rightarrow 0] 
}
\label{4.15}
\end{equation}

\noindent where, using $\Delta_3 = \Delta_1 + \Delta_{\MM} + \Delta_2$, we have

\begin{multline}
\label{w.numerator.4D}
w^{\, num} [{\bf V}, {\bf W}, {\bf \Delta}, \epsilon_1, \epsilon_2, R \rightarrow 0] 
= 
\\
\prod_{\ws \in V^{\, 1}} \ll \Delta_1     + A^+_{\ws, W^{\, 1}} \et - L^+_{\ws, V^{\, 1}} \eo \rr 
\prod_{\bs \in W^{\, 1}} \ll \Delta_1     - A^+_{\bs, V^{\, 1}} \et + L^+_{\bs, W^{\, 1}} \eo \rr 
\\
\times   
\prod_{\ws \in V^{\, 2}} \ll \Delta_2     + A^+_{\ws, W^{\, 2}} \et - L^+_{\ws, V^{\, 2}} \eo \rr 
\prod_{\bs \in W^{\, 2}} \ll \Delta_2     - A^+_{\bs, V^{\, 2}} \et + L^+_{\bs, W^{\, 2}} \eo \rr 
\\
\times   
\prod_{\ws \in V^{\, 2}} \ll \Delta_{\MM} - A^+_{\ws, W^{\, 1}} \et + L^+_{\ws, V^{\, 2}} \eo \rr 
\prod_{\bs \in W^{\, 1}} \ll \Delta_{\MM} + A^+_{\bs, V^{\, 2}} \et - L^+_{\bs, W^{\, 1}} \eo \rr 
\\
\times 
\prod_{\ws \in V^{\, 1}} \ll \Delta_3     + A^+_{\ws, W^{\, 2}} \et - L^+_{\ws, V^{\, 1}} \eo \rr 
\prod_{\bs \in W^{\, 2}} \ll \Delta_3     - A^+_{\bs, V^{\, 1}} \et + L^+_{\bs, W^{\, 2}} \eo \rr 
\end{multline}

\noindent and

\begin{multline}
\label{w.denominator.4D}
w_{\, den} [{\bf V}, {\bf W}, {\bf \Delta}, \epsilon_1, \epsilon_2, R \rightarrow 0] 
=   
\\
\prod_{\ws \in V^{\, 1}} \ll                        - A^{++}_{\ws, V^{\, 1}} \et + L^{  }_{\ws, V^{\, 1}} \eo \rr 
\prod_{\ws \in V^{\, 2}} \ll                        - A^{++}_{\ws, V^{\, 2}} \et + L^{  }_{\ws, V^{\, 2}} \eo \rr 
\\
\times   
\prod_{\bs \in W^{\, 1}} \ll                        - A^{  }_{\bs, W^{\, 1}} \et + L^{++}_{\bs, W^{\, 1}} \eo \rr 
\prod_{\bs \in W^{\, 2}} \ll                        - A^{  }_{\bs, W^{\, 2}} \et + L^{++}_{\bs, W^{\, 2}} \eo \rr 
\\
\times   
\prod_{\ws \in V^{\, 1}} \ll \Delta_1 + \Delta_{\MM} + A^{++}_{\ws, V^{\, 2}} \et - L^{  }_{\ws, V^{\, 1}} \eo \rr 
\prod_{\ws \in V^{\, 2}} \ll \Delta_1 + \Delta_{\MM} - A^{  }_{\ws, V^{\, 1}} \et + L^{++}_{\ws, V^{\, 2}} \eo \rr 
\\
\times   
\prod_{\bs \in W^{\, 1}} \ll \Delta_2 + \Delta_{\MM} + A^{  }_{\bs, W^{\, 2}} \et - L^{++}_{\bs, W^{\, 1}} \eo \rr 
\prod_{\bs \in W^{\, 2}} \ll \Delta_2 + \Delta_{\MM} - A^{++}_{\bs, W^{\, 1}} \et + L^{  }_{\bs, W^{\, 2}} \eo \rr 
\end{multline}

\section{The building block of the 4D $U(2)$ quiver instanton partition function}
\label{section.05.nekrasov.partition.functions}
{\it 
We recall the normalised contribution of the bifundamental hypermultiplet which acts as a building block 
of the instanton partition function.
}

In the notation of Equation {\bf 6} and Section {\bf 2} of \cite{bershtein.foda}, the normalised bifundamental 
partition function 
$\cZ^{\, 4D}_{building.block}$ is 
\begin{equation}
\label{z.bb}
\cZ^{\, 4D}_{building.block}  \ll {\bf a}, {\bf V^{\prime}} \ | \ \mu \ | \ {\bf b}, {\bf W^{\prime}} \rr = 
\frac{
z^{\, num} \ll {\bf a}, {\bf V^{\prime}} \ | \ \mu \ | \ {\bf b}, {\bf W^{\prime}} \rr
}
{
z_{\, den} \ll {\bf a}, {\bf V^{\prime}} \ |           \ {\bf b}, {\bf W^{\prime}} \rr
} 
\end{equation}

\noindent where ${\bf a} = \{a, -a\}$, ${\bf b} = \{b, -b\}$, $a, b$ and $\mu$ are linear combinations 
of $\eo$ and $\et$, as will be explained in 
Section {\bf \ref{section.08.restricted.instanton.partition.functions}} below, and we use 
${\bf V^{\prime}}$ and 
${\bf W^{\prime}}$ for partition pairs that we will relate in 
Section {\bf \ref{section.07.identification.denominators}} to the pairs that appear in the 4D basic web. 
We refer to \cite{bershtein.foda} for brief explanations of the parameters that appear in 
$\cZ^{\, 4D}_{building.block}$. Defining

\begin{equation}
\alpha_0 = \frac12 \ll \eo + \et \rr
\end{equation}

\noindent the numerator $z^{\, num}$, as given in Equation {\bf 9} of \cite{bershtein.foda}, is

\begin{multline}
\label{z.numerator}
z^{\, num} \ll {\bf a}, {\bf V^{\prime}}\  | \ \mu \ | \ {\bf b}, {\bf W^{\prime}} \rr  = 
\\
\prod_{\ws \in V^{\, 1 \, \prime}} \ll [  a - b - \mu + \alpha_0] + A^+_{\ws, V^{\, 1 \, \prime}} \eo 
                                                                  - L^+_{\ws, W^{\, 1 \, \prime}} \et \rr
\prod_{\bs \in W^{\, 1 \, \prime}} \ll [  a - b - \mu + \alpha_0] - A^+_{\bs, W^{\, 1 \, \prime}} \eo 
                                                                  + L^+_{\bs, V^{\, 1 \, \prime}} \et \rr
\\
\times 
\prod_{\ws \in V^{\, 2 \, \prime}} \ll [- a + b - \mu + \alpha_0] + A^+_{\ws, V^{\, 2 \, \prime}} \eo 
                                                                  - L^+_{\ws, W^{\, 2 \, \prime}} \et \rr
\prod_{\bs \in W^{\, 2 \, \prime}} \ll [- a + b - \mu + \alpha_0] - A^+_{\bs, W^{\, 2 \, \prime}} \eo 
                                                                  + L^+_{\bs, V^{\, 2 \, \prime}} \et \rr
\\
\times 
\prod_{\ws \in V^{\, 2 \, \prime}} \ll [- a - b - \mu + \alpha_0] + A^+_{\ws, V^{\, 2 \, \prime}} \eo 
                                                                  - L^+_{\ws, W^{\, 1 \, \prime}} \et \rr
\prod_{\bs \in W^{\, 1 \, \prime}} \ll [- a - b - \mu + \alpha_0] - A^+_{\bs, W^{\, 1 \, \prime}} \eo 
                                                                  + L^+_{\bs, V^{\, 2 \, \prime}} \et \rr
\\
\times
\prod_{\ws \in V^{\, 1 \, \prime}} \ll [  a + b - \mu + \alpha_0] + A^+_{\ws, V^{\, 1 \, \prime}} \eo 
                                                                  - L^+_{\ws, W^{\, 2 \, \prime}} \et \rr
\prod_{\bs \in W^{\, 2 \, \prime}} \ll [  a + b - \mu + \alpha_0] - A^+_{\bs, W^{\, 2 \, \prime}} \eo 
                                                                  + L^+_{\bs, V^{\, 1 \, \prime}} \et \rr
\end{multline}

The denominator $z_{\, den}$, as given in Equation {\bf 7} of \cite{bershtein.foda}, is

\begin{equation}
\label{z.denominator}
z_{\, den} \ll {\bf a}, {\bf V^{\prime}} \ | \ {\bf b}, {\bf W^{\prime}} \rr
= 
\ll 
z^{\, num} \ll {\bf a}, {\bf V^{\prime}} \ | \  0 \ | \ {\bf a}, {\bf V^{\prime}} \rr \ 
z^{\, num} \ll {\bf b}, {\bf W^{\prime}} \ | \  0 \ | \ {\bf b}, {\bf W^{\prime}} \rr \ 
\rr^{\frac{1}{2}}
\end{equation}

\section{Identification of $\cW_{\bf \, V \, W \, \Delta}^{\, norm}$ and $\cZ^{\, 4D}_{building.block}$. The numerators}
\label{section.06.identification.numerators}

Comparing Equation {\bf \ref{w.numerator.4D}} and Equation {\bf \ref{z.numerator}}, we find that if 
we set  
$V^{\, i \, \prime} = V^{\, i \, \intercal}$, 
$W^{\, i \, \prime} = W^{\, i \, \intercal}$, and multiply each factor by $-1$, which is possible 
since the number of factors is even by construction, we obtain

\begin{eqnarray}
\label{numerator.nekrasov.02}
&        & 
\\
&        & z^{\, num} \ll {\bf a}, {\bf V}\,  | \ \mu \ | \, {\bf b}, {\bf W} \rr =
\nonumber
\\
&        &
\prod_{\ws \in V^{\, 1}} \ll [- a + b + \mu - \alpha_0] + A^+_{\ws, W^{\, 1}} \et - L^+_{\ws, V^{\, 1}} \eo \rr
\prod_{\bs \in W^{\, 1}} \ll [- a + b + \mu - \alpha_0] - A^+_{\bs, V^{\, 1}} \et + L^+_{\bs, W^{\, 1}} \eo \rr
\nonumber
\\
& \times &
\prod_{\ws \in V^{\, 2}} \ll [  a - b + \mu - \alpha_0] + A^+_{\ws, W^{\, 2}} \et - L^+_{\ws, V^{\, 2}} \eo \rr
\prod_{\bs \in W^{\, 2}} \ll [  a - b + \mu - \alpha_0] - A^+_{\bs, V^{\, 2}} \et + L^+_{\bs, W^{\, 2}} \eo \rr
\nonumber
\\
& \times &
\prod_{\ws \in V^{\, 2}} \ll [  a + b + \mu - \alpha_0] + A^+_{\ws, W^{\, 1}} \et - L^+_{\ws, V^{\, 2}} \eo \rr
\prod_{\bs \in W^{\, 1}} \ll [  a + b + \mu - \alpha_0] - A^+_{\bs, V^{\, 2}} \et + L^+_{\bs, W^{\, 1}} \eo \rr
\nonumber
\\
& \times &
\prod_{\ws \in V^{\, 1}} \ll [- a - b + \mu - \alpha_0] + A^+_{\ws, W^{\, 2}} \et - L^+_{\ws, V^{\, 1}} \eo \rr
\prod_{\bs \in W^{\, 2}} \ll [- a - b + \mu - \alpha_0] - A^+_{\bs, V^{\, 1}} \et + L^+_{\bs, W^{\, 2}} \eo \rr
\nonumber
\end{eqnarray}

\noindent which leads to the identification

\begin{equation}
\label{identification.12}
\boxed{
\Delta_1     =          -   a + b + \mu - \alpha_0, \quad
\Delta_2     = \phantom{-}  a - b + \mu - \alpha_0, \quad
\Delta_{\MM} =          -   a - b - \mu + \alpha_0
}
\end{equation}

\section{Identification of $\cW_{\bf \, V \, W \, \Delta}^{\, norm}$ and $\cZ^{\, 4D}_{building.block}$. The denominators}
\label{section.07.identification.denominators}

Using the identification of parameters obtained in Equation {\bf \ref{identification.12}} in
$w_{\, den} [{\bf V, \, W, \, \Delta}]$ 
and $z_{\, den}$ as given in Equations {\bf \ref{w.denominator.4D}} and {\bf \ref{z.denominator}}, 
it is clear that these two functions are not the same. However, what matters is not the denominator 
of s single factor, but the product of all denominators, as we explain below.

The denominator $w_{\, den} [{\bf V, \, W, \, \Delta}]$ is a natural object, as we can see in the 
derivation in Section {\bf \ref{section.03.5d.web}}. 
On the other hand, the denominator $z_{\, den}$ was obtained in \cite{bershtein.foda} by taking the 
full denominator that appears in expressions for the 4D $U(1)$ linear and cyclic quiver instanton 
partition functions and factoring that into denominators for the contributions of the bifundamental 
hypermultiplets.  Such a factorisation is not unique and any factorisation is allowed for as long as 
the product of all factors is equal to the full denominator of the original expression. 

In this work, to identify $\cW^{\, 4D}$ and $\cZ^{\, 4D}_{building.block}$, we need 
a factor $\cF$, such that the product of all normalisation factors that appear in a conformal block 
is equal to 1. Consider the abbreviations
\begin{eqnarray}
\cA_{V^{ij}} [x] = \cA_{V^i, V^j} [x] & = & \prod_{\ws \in V^i} \ll x + A^{++}_{\ws, V^i} \et
- L^{  }_{\ws, V^j} \eo \rr,
\quad
\cA_{\emptyset, \emptyset} [x] = 1,
\\
\cL_{W^{ij}} [x] = \cL_{W^i, W^j} [x] & = & \prod_{\bs \in W^j} \ll x - A^{  }_{\bs, W^j} \et
+ L^{++}_{\bs, W^i} \eo \rr,
\quad
\cL_{\emptyset, \emptyset} [x] = 1,
\nonumber
\\
\cH_{Y^{ij}} [x] & = & \cA_{Y^{ij}} [x] \, \cL_{Y^{ij}} [x]
\nonumber
\end{eqnarray}
In this notation, $z_{\, den}$ and $w_{\, den}$ are 
\begin{multline}
z_{\, den} \ll {\bf a}, {\bf V} \ | \ 0 \ | \ {\bf b}, {\bf W} \rr =
\\
\ll
\cH_{V^{11}} [0] \, \cH_{V^{12}} [2a] \, \cH_{V^{12}} [-2a] \, \cH_{V^{22}} [0]
\cH_{W^{11}} [0] \, \cH_{W^{12}} [2b] \, \cH_{W^{12}} [-2b] \, \cH_{W^{22}} [0]
\rr^{\frac12}
\end{multline}

\noindent and 
\begin{multline}
w_{\, den} [{\bf V, \, W, \, \Delta}, \epsilon_1, \epsilon_2, R \rightarrow 0] = 
w_{\, den} \ll {\bf a}, {\bf V} \, | \, {\bf b}, {\bf W} \rr =
\\
\ll - \rr^{\, | V^{\, 1} | + | V^{\, 2} | + | W^{\, 1} | + | W^{\, 2} |}
\cL_{V^{11}} [  0] \,
\cL_{V^{22}} [  0] \,
\cH_{V^{12}} [-2a] \,
\cH_{W^{12}} [ 2b] \,
\cA_{W^{11}} [  0] \,
\cA_{W^{22}} [  0]
\end{multline}

Now consider the factor

\begin{equation}
\cF \ll {\bf a}, {\bf V} \, | \, {\bf b}, {\bf W} \rr
=
\frac{
z_{\, den} \left[ {\bf a}, {\bf V} \, | \, {\bf b}, {\bf W} \right]
}{
w_{\, den} \left[ {\bf a}, {\bf V} \, | \, {\bf b}, {\bf W} \right]
}
\end{equation}

\noindent and define 

\begin{equation}
z_{\, den}^{\prime} = {\cF}^{-1} \, z_{\, den} = w_{\, den}, \quad
\cZ_{building.block }^{\prime} =  \cF       \, \cZ_{building.block}
\end{equation}

\noindent $\cZ^{\, \prime}_{building.block}$ is constructed such that 
{\bf 1.} It has the same numerator as $\cZ^{\, 4D}_{building.block}$, which 
is the same as that of $\cW^{\, norm}_{\bf V W \Delta}$, 
when we choose the parameters as in Equation {\bf \ref{identification.12}} and
{\bf 2.} It has the same denominator as $\cW^{\, norm}_{\bf V W \Delta}$, also 
when we choose the parameters as in Equation {\bf \ref{identification.12}}.

Since the denominator of $\cZ^{\, \prime}_{building.block}$ is not manifestly 
the same as that of $\cZ^{\,     4D}_{building.block}$, we need to show that 
gluing copies    of $\cZ^{\, \prime}_{building.block}$ to build a topological 
partition function, leads to the same result obtained by gluing copies of the 
original            $\cZ^{         }_{building.block}$. $\cF$ can be written 
in a simpler form as follows,

\begin{equation}
\ll \cF \ll {\bf a}, {\bf V} \, | \, {\bf b}, {\bf W} \rr \rr^{\, 2} 
\\
=
F^{\,  left}_{\bf V} [a] \,
F^{\, right}_{\bf W} [b]
\end{equation}

\noindent where 

\begin{eqnarray}
F^{\, left}_{\bf V} [a] & = &
\ll - \rr^{\, | V^{\, 1} | + | V^{\, 2} | }
\ll
\frac{
\cA_{V^{11}} [0] \,
\cA_{V^{22}} [0]
}{
\cL_{V^{11}} [0] \,
\cL_{V^{22}} [0]
}
\rr \, 
\ll
\frac{
\cH_{V^{12}} [ 2a]
}{
\cH_{V^{12}} [-2a]
}
\rr 
\\
F^{\, right}_{\bf W} [b] & = &
\ll - \rr^{\, | W^{\, 1} | + | W^{\, 2} |}
\ll
\frac{
\cH_{W^{12}} [-2b]
}{
\cH_{W^{12}} [ 2b]
}
\rr \, 
\ll
\frac{
\cL_{W^{11}} [0] \,
\cL_{W^{22}} [0]
}{
\cA_{W^{11}} [0] \,
\cA_{W^{22}} [0]
}
\rr
\end{eqnarray}

\noindent $F^{\, right}$ and $F^{\, left }$ satisfy the obvious properties 

\begin{equation}
\label{property.00}
F^{\, right}_{\bf \emptyset} [x] =
F^{\, left }_{\bf \emptyset} [x] = 1
\end{equation}

\noindent and

\begin{equation}
\label{property.01}
F^{\, right}_{\bf Y} [x] \, F^{\, left }_{\bf Y} [x] = 1
\end{equation}

The physical objects that we are interested in are the conformal blocks which are constructed 
by gluing copies of $\cZ_{building.block}$ \cite{bershtein.foda}. We need to show that gluing 
copies of $\cZ_{building.block}^{\prime}$ leads to the same result, which will be the case if 
products of the normalisation factors trivialise. This will follow directly from Equations 
{\bf \ref{property.00}} and {\bf \ref{property.01}}. There are two cases to consider, the 
linear conformal block case and the cyclic conformal block case.

\subsubsection{Linear conformal blocks}
\label{linear.conformal.blocks}
Consider the linear conformal block obtained by gluing $n$ copies of $\cZ_{building.block}$, 
that is $\cZ_{building.block.1}$, $\cZ_{building.block.2}$, $\cdots$, $\cZ_{building.block.n}$, 
sequentially. Using copies of $\cZ_{building.block}^{\prime}$, we obtain the same result as 
using copies of $\cZ_{building.block}$ up to a factor 

\begin{equation}
F^{\, left}_{\bf  \emptyset} [x_0    ] \, F^{\, right}_{\bf    Y^{\, 1}} [x_1] 
F^{\, left}_{\bf     Y^{\, 1}} [x_1    ] \, F^{\, right}_{\bf    Y^{\, 2}} [x_2] 
\cdots
F^{\, left}_{\bf Y^{n-1}} [x_{n-1}] \, F^{\, right}_{\bf \emptyset} [x_n] 
= 1
\end{equation}

\subsubsection{Cyclic conformal blocks}
\label{cyclic.conformal.blocks}

\begin{equation}
F^{\, left}_{\bf  Y^0} [x_0] \, F^{\, right}_{\bf    Y^{\, 1}} [x_1]
F^{\, left}_{\bf  Y^{\, 1}} [x_1] \, F^{\, right}_{\bf    Y^{\, 2}} [x_2]
\cdots
F^{\, left}_{\bf Y^{n-1}} [x_{n-1}] \, F^{\, right}_{\bf Y^0} [x_0]
= 1
\end{equation}

We conclude that $\cZ_{building.block}^{\prime}$ leads to the same conformal blocks as $\cZ_{building.block}$.  

\subsubsection{The denominator $z_{\, den}^{\prime}$ and the Burge conditions}
In \cite{alkalaev.belavin, bershtein.foda}, it was shown that for the choice of parameters that 
leads to Virasoro $A$-series minimal conformal blocks, the denominator $z_{\, den}$ will contain non-physical 
zeros, unless we restrict the partition pairs, that $\cZ^{\, 4D, \, min}_{building.block}$ depends on, to obey 
Burge conditions. These conditions were derived in \cite{bershtein.foda} using $z_{\, den}$ rather than 
$z_{\, den}^{\prime}$. Using $z_{\, den}^{\prime}$ leads to the same conditions, since the product of all 
$z_{\, den}^{\prime}$ is the same as the product of all $z_{\, den}$ that show up in the conformal block
\footnote{\,
In Section {\bf \ref{section.10.burge.pairs}}, we outline the derivation of the Burge conditions from 
$z_{\, den}^{\prime}$.}.

\section{Restricted instanton partition functions for $\cM^{\, p, \, \pp, \, \cH}$. The parameters}
\label{section.08.restricted.instanton.partition.functions}

{\it 
We recall the choice of parameters such that $\cZ^{\, 4D}_{building.block}$ reduces to 
$\cZ^{\, 4D, \, min}_{building.block}$ which is the building block of Virasoro $A$-series minimal 
model conformal blocks times Heisenberg factors.
}

\subsection{AGT parameterisation of minimal models}
\label{parameterisation.minimal.models}

A minimal model $\cM^{\, p, \, \pp}$, based on a Virasoro algebra $\cV^{\, p, \, \pp}$, characterised 
by a central charge $c_{p, \, \pp} < 1$, that we parameterise as

\begin{equation}
\label{central.charge.minimal}
c_{p, \, \pp} = 1 - 6 \ll a_{p, \, \pp} - \frac{1}{a_{p, \, \pp}} \rr^{\, 2},
\quad
a_{p, \, \pp} = \ll \frac{\pp}{p} \rr^{\frac12},
\end{equation}

\noindent where $\{p, \, \pp\}$ are co-prime integers that satisfy $0 < p < \pp$. In the Coulomb 
gas approach to computing conformal blocks in minimal models with $c_{p, \, \pp} < 1$ 
\cite{nienhuis, dotsenko.fateev}, the screening charges $\{\apos, \aneg\}$, and the background 
charge $\alpha_{back.ground}$, satisfy 

\begin{equation}
\label{minimal.model.charges}
\boxed{
\apos =             a_{p, \, \pp}, 
\quad
\aneg = -  \frac{1}{a_{p, \, \pp}},
\quad
\alpha_{back.ground} = - 2 \alpha_0, 
\quad 
\alpha_0  = \frac12 \ll \apos + \aneg \rr 
}
\end{equation}

\noindent The AGT parameterisation of $\cM^{\, p, \, \pp, \, \cH}$ is obtained by choosing 

\begin{equation}
\label{neg.0.pos}
\boxed{
\eo < 0 < \et, 
\quad
\eo = \aneg, 
\quad
\et = \apos 
}
\end{equation}

\noindent so that $\aneg < 0 < \apos$. Since we focus on $\cM^{\, p, \, \pp, \, \cH}$, 
we work in terms of $\{\aneg, \apos\}$ instead of $\{ \eo, \et\}$.

\subsection{Two sets of charges in minimal models}
\label{charge.content}
We consider two types of charges that, in Coulomb gas terms, are expressed in terms of 
the screening charges $\{\apos, \aneg\}$. 
{\bf 1.} The charge $a_{r, s}$ of the highest weight $|a_{r, s} \rangle$ of the irreducible
highest weight representation $\cH^{\, p, \, \pp}_{r, s}$, and
{\bf 2.} The charge $\mu_{r, s}$ of the vertex operator ${\cO}_{\mu}$ that intertwines 
two highest weight ireducible representations 
$\cH^{\, p, \, \pp}_{r_1, s_1}$  and 
$\cH^{\, p, \, \pp}_{r_2, s_2}$.
These charges are parameterised in terms of $\apos$ and $\aneg$ as follows 

\begin{equation}
\label{parameters.03}
\boxed{
  a_{r, s} = - \frac{r}{2} \, \apos - \frac{s}{2} \, \aneg,            \quad 
\mu_{r, s} = - \frac{r}{2} \, \apos - \frac{s}{2} \, \aneg + \alpha_0, \quad 
1 \leq r \leq   p - 1,
1 \leq s \leq \pp - 1
}
\end{equation}

\section{From gauge theory parameters to minimal model parameters}
\label{section.09.from.gauge.theory.to.minimal.model}
{\it We compare the parameters of $\cW^{\, 4D}$ and the parameters of $\cZ^{\, 4D}_{building.block}$.}

We set

\begin{equation}
\label{notation.02.01}
a = - \ll \frac{r_a}{2}        \rr \, \apos - \ll \frac{s_a}{2}     \rr \, \aneg, 
\quad
r_a \in \{1, 2, \cdots,    p - 1\}, 
\quad
s_a \in \{1, 2, \cdots,  \pp - 1\}, 
\end{equation}

\begin{equation}
\label{notation.02.02}
b    = - \ll \frac{r_b}{2}     \rr \, \apos - \ll \frac{s_b}{2}     \rr \, \aneg, 
\quad
r_b \in \{1, 2, \cdots,   p - 1\},
\quad
s_b \in \{1, 2, \cdots, \pp - 1\},
\end{equation}

\begin{equation}
\label{notation.02.03}
\mu  = - \ll \frac{r_{\mu}}{2} \rr \, \apos - \ll \frac{s_{\mu}}{2} \rr \, \aneg + \alpha_0,
\quad
r_{\mu} \in \{1, 2, \cdots,   p - 1\}, 
\quad
s_{\mu} \in \{1, 2, \cdots, \pp - 1\}
\end{equation}

\subsection{The fusion rules}
For completeness, let us mention the fusion rules. In the notation

\begin{equation}
m_i = r_i - 1, 
\quad 
n_i = s_i - 1
\end{equation}

\noindent the fusion rules take the simple form

\begin{equation}
\label{fusion.rules}
m_a + m_b + m_{\mu} = 0  \ \textit{mod}  \ 2, \quad
n_a + n_b + n_{\mu} = 0  \ \textit{mod}  \ 2,
\end{equation}

\noindent where the triple $\{m_a, m_b, m_{\mu}\}$ satisfies the triangular conditions

\begin{equation}
\label{triangular.conditions}
m_a + m_b     \geq m_{\mu}, \quad 
m_b + m_{\mu} \geq m_a,     \quad 
m_{\mu} + m_a \geq m_b
\end{equation}

\noindent with analogous conditions for the triple $\{n_a, n_b, n_{\mu}\}$. 

\section{Restricted instanton partition functions for $\cM^{\, p, \, \pp, \, \cH}$. 
The partition pairs}
\label{section.10.burge.pairs}
{\it We write the denominator 
$w_{\, den} [{\bf V, \, W, \, \Delta}, \eo,   \et,   g \rightarrow 0]$ as
$w_{\, den} [{\bf V, \, W, \, \Delta}, \apos, \aneg                 ]$, and, following 
\cite{bershtein.foda}, we check the conditions required so that that it has no zeros.}

Using $\apos$ and $\aneg$, let us write 
$w_{\, den} [{\bf V, \, W, \, \Delta}, \eo,   \et, R \rightarrow 0]$ as
$w_{\, den} [{\bf V, \, W, \, \Delta}, \apos, \aneg]$, that is 

\begin{eqnarray}
\label{denominator.4D.02}
w_{\, den} [{\bf V, \, W, \, \Delta}, \apos, \aneg]
& = &
\prod_{\ws \in V^{\, 1}} \ll                                 A^{++}_{\ws, V^{\, 1}} \, \apos - L^{  }_{\ws, V^{\, 1}} \, \aneg \rr
\prod_{\ws \in V^{\, 2}} \ll                                 A^{++}_{\ws, V^{\, 2}} \, \apos - L^{  }_{\ws, V^{\, 2}} \, \aneg \rr
\nonumber
\\
& \times &
\prod_{\bs \in W^{\, 1}} \ll                                 A^{  }_{\bs, W^{\, 1}} \, \apos - L^{++}_{\bs, W^{\, 1}} \, \aneg \rr
\prod_{\bs \in W^{\, 2}} \ll                                 A^{  }_{\bs, W^{\, 2}} \, \apos - L^{++}_{\bs, W^{\, 2}} \, \aneg \rr
\nonumber
\\
& \times & 
\prod_{\ws \in V^{\, 1}} \ll r_a \, \apos + s_a \, \aneg  +  A^{++}_{\ws, V^{\, 2}} \, \apos - L^{  }_{\ws, V^{\, 1}} \, \aneg \rr
\nonumber
\\
& \times &
\prod_{\ws \in V^{\, 2}} \ll r_a \, \apos + s_a \, \aneg  -  A^{  }_{\ws, V^{\, 1}} \, \apos + L^{++}_{\ws, V^{\, 2}} \, \aneg \rr
\nonumber
\\
& \times & 
\prod_{\bs \in W^{\, 1}} \ll r_b \, \apos + s_b \, \aneg  +  A^{  }_{\bs, W^{\, 2}} \, \apos - L^{++}_{\bs, W^{\, 1}} \, \aneg \rr
\nonumber
\\
& \times &
\prod_{\bs \in W^{\, 2}} \ll r_b \, \apos + s_b \, \aneg  -  A^{++}_{\bs, W^{\, 1}} \, \apos + L^{  }_{\bs, W^{\, 2}} \, \aneg \rr
\end{eqnarray}

\noindent where we have used 

\begin{eqnarray}
-2a & = & r_a \, \apos + s_a \, \aneg, \quad r = 1, 2, \cdots, 
\\
-2b & = & r_b \, \apos + s_b \, \aneg, \quad s = 1, 2, \cdots
\end{eqnarray}

Consider the denominator $w_{\, den} [{\bf V, \, W, \, \Delta}, \apos, \aneg]$ in 
Equation {\bf \ref{denominator.4D.02}}, 
on a product by product basis. We need to check the conditions under which any of these products has a zero, then
find the restriction that are necessary and sufficient to remove these zeros. The reasoning that we use to obtain
these conditions is the same as that in \cite{bershtein.foda}. There are eight products to consider.

\subsection{The initial four products}  In each of the initial four products, the product is over the cells inside 
a single diagram, thus the arm length $A$ and the leg length $L$ in each of these factors is non-negative. Since 
$\aneg < 0 < \apos$, and there is a term $\alpha_0 > 0$ in each factor, the minimal value of each of these factors 
is greater than zero. Thus there can be no zeros from these factors. 
To consider the remaining four factors, we require some preparation.

\subsection{Two zero-conditions}
\label{to.vanish}
Following \cite{bershtein.foda}, we note that, since $\aneg < 0 < \apos$, any factor of the type that 
appears in Equation {\bf \ref{denominator.4D.02}} has a zero when an equation of type

\begin{equation}
\label{zeros.example.01}
C_{+} \, \apos + C_{-} \, \aneg = 0,
\end{equation}

\noindent where $C_+$, $C_-$ $\in \ZZ$, is satisfied. Since $p$ and $\pp$ are coprime, $\aneg$ 
and $\apos$ are $\not \in \QQ$, the condition in Equation {\bf \ref{zeros.example.01}} is 
equivalent to the two conditions

\begin{equation}
\label{zeros.example.02}
C_+ =  c   \ p, 
\quad
C_- =  c \ \pp
\end{equation}

\noindent are satisfied, where $c$ is a proportionality constant that needs to be determined. 

\subsection{From two zero-conditions to one zero-condition} 
\label{from.2.to.1}

Consider the two conditions

\begin{equation}
\label{conditions.example.01}
\boxed{
- A^{ }_{\square, i}   =  A^{\prime} \geq 0,
\quad 
  L^{ }_{\square, j}   =  L^{\prime} \geq 0
}
\end{equation}

\noindent which are satisfied if $i \neq j$, $\square \not \in Y^i$, and $\square \in Y^j$.
If $\square$ is in row-$\rrho$ and column-$\ssigma$ in $Y^j$, then the second condition in 
(\ref{conditions.example.01}) implies that there is a cell $\boxplus \in Y^{\, 1}$, strictly 
below $\square$, with coordinates $\{\rrho + L^{\prime}, \ssigma\}$, such that there are 
no cells strictly below $\boxplus$. Since there may, or may not, be cells to the right of 
$\boxplus$, row-$(\rrho + L^{\prime})$ in $Y^j$ has length {\it at least} $\ssigma$,

\begin{equation}
\label{conditions.example.02}
y^{\, j}_{\, \rrho + L^{\prime}} \geq \ssigma
\end{equation}

\noindent From the definition of $A_{\square, \, i}$, we write the first condition in 
(\ref{conditions.example.01}) as 
$- A_{\square, \, i}   = A^{\prime} = \ssigma - y^{\, i}_{\, \rrho}$,
that is, 
$\ssigma = A^{\prime} + y^{\, i}_{\, \rrho}$, 
and using (\ref{conditions.example.02}), we obtain 
$y^{\, j}_{\, \rrho + L^{\prime}} \geq A^{\prime} + y^{\, i}_{\, \rrho}$,
which we choose to write as

\begin{equation}
\label{conditions.example.04}
y^{\, j}_{\, \rrho + L^{\prime}} - y^{\, i}_{\, \rrho} \geq A^{\prime}
\end{equation}

\noindent The condition in Equation {\bf \ref{conditions.example.04}} is equivalent to 
the two conditions in Equation {\bf \ref{conditions.example.01}}. 

\subsection{One non-zero condition}
\label{1.not.to.vanish}
Consider a function $f_{Y^{\, i}, Y^{\, j}}$, of a pair of Young diagrams $Y^{\, i}$ and $Y^{\, j}$, 
$i \neq j$, such that 
$f_{Y^{\, i}, Y^{\, j}} = 0$, {\it if and only if} (\ref{conditions.example.04}) is satisfied. This 
implies that $f_{Y^{\, i}, Y^{\, j}} \neq 0$, {\it if and only if} $Y^{\, i}$ and $Y^{\, j}$ satisfies 
the complementary condition

\begin{equation}
\label{conditions.example.05}
y^{\, j}_{\, \rrho + L^{\prime}} - y^{\, i}_{\, \rrho} < A^{\prime}
\end{equation}

\noindent which we choose to write as 

\begin{equation}
\label{conditions.example.06}
\boxed{
y^{\, i}_{\, \rrho} - y^{\, j}_{\, \rrho + L^{\prime}} \geq 1 - A^{\prime}
}
\end{equation}

\subsubsection{Remark} 
In the sequel, we refer 
to Equation {\bf \ref{conditions.example.01}} as {\it \lq       zero-conditions\rq\,}, and 
to Equation {\bf \ref{conditions.example.06}} as {\it \lq a non-zero-condition\rq\,}. 

Next we consider the latter four products on the right hand side of Equation 
{\bf \ref{denominator.4D.02}}.

\subsection{The first product} 
\label{den.12}

\begin{equation}
\prod_{\ws \in V^{\, 1}} \ll  r_a \, \apos + 
                         s_a \, \aneg + A^{++}_{\ws, \, V^{\, 2}} \, \apos 
                                      - L^{  }_{\ws, \, V^{\, 1}} \, \aneg \rr
\end{equation}

\noindent vanishes if any factor satisfies 

\begin{equation}
\label{zeros.12}
\ll  r_a + A^{++}_{\square, \, V^{\, 2}} \rr \, \apos +
\ll  s_a - L^{  }_{\square, \, V^{\, 1}} \rr \, \aneg = 0
\end{equation}
\noindent which leads to the conditions
\begin{equation}
\label{two.zero.conditions.12}
- A^{ }_{\ws, \, V^{\, 2}} = r_a + 1 + c \,    p, \quad 
  L^{ }_{\ws, \, V^{\, 1}} = s_a     + c \,  \pp 
\end{equation}

\noindent Since $\ws \in V^{\, 1}$, $L_{\ws, V^{\, 1}} \geq 0$. Given that $s$ and $\pp$ are 
non-zero positive integers, 
the second equation in Equation {\bf \ref{two.zero.conditions.12}} admits a solution 
only if $c = 0, 1, \cdots$ 
The first  equation in Equation {\bf \ref{two.zero.conditions.12}} admits a solution 
if $\ws \not \in V^{\, 2}$.

\subsubsection{From two zero-conditions to one non-zero-condition} 
\label{translating.01}
Following paragraphs {\bf \ref{from.2.to.1}} and {\bf \ref{1.not.to.vanish}}, the two 
zero-conditions in (\ref{two.zero.conditions.12}) are equivalent to one non-zero-condition, 

\begin{equation}
\label{one.non.zero.condition.12}
V^{\, 2}_{\rrho} - V^{\, 1}_{\rrho + s + c \, \pp} \geq - r - c \, p 
\end{equation}

\subsubsection{The stronger condition} 
\label{the.stronger.condition.01}
Equation {\bf \ref{one.non.zero.condition.12}} is the statement that to eliminate the zeros, 
we want 
$V^{\, 2}_{\rrho} - V^{\, 1}_{\rrho + s + c \, \pp} \geq - r - c \, p$, 
where $c = \{0, 1, \cdots\}$
Since the row-lengths of a partition are by definition weakly decreasing, and 
$c = \{0, 1, \cdots\}$, this is the case if 
$V^{\, 2}_{\rrho} - V^{\, 1}_{\rrho + s} \geq - r - c \, p$, 
which is the case if 
$V^{\, 2}_{\rrho} - V^{\, 1}_{\rrho + s} \geq - r$. 
Thus, we should set $c=0$, and obtain 

\begin{equation}
\label{weak.Burge.condition.02}
V^{\, 2}_{\rrho} - V^{\, 1}_{\rrho + s} \geq - r 
\end{equation}

\subsection{The second product} 
\label{den.21}

\begin{equation}
\prod_{\ws \in V^{\, 2}} \ll  r_a \, \apos + s_a \, \aneg  -  A^{  }_{\ws, V^{\, 1}} \, \apos 
                                                           +  L^{++}_{\ws, V^{\, 2}} \, \aneg \rr
\end{equation}

\noindent vanishes if any factor satisfies 
\begin{equation}
\label{zeros.21}
\ll r_a - A^{  }_{\ws, V^{\, 1}} \rr \, \apos 
+ 
\ll s_a + L^{++}_{\ws, V^{\, 2}} \rr \, \aneg = 0
\end{equation}
\noindent which leads to the conditions
\begin{equation}
\label{two.zero.conditions.21}
- A^{ }_{\ws, V^{\, 1}} = - r_a     + c \,   p, 
\quad
  L^{ }_{\ws, V^{\, 2}} = - s_a - 1 + c \, \pp,
\end{equation}

\noindent Since $\ws \in V^{\, 2}$, $L_{\ws, V^{\, 2}} \geq 0$. Given that $s_a$ and $\pp$ are 
non-zero positive integers,
the second equation in Equation {\bf \ref{two.zero.conditions.21}} admits a solution 
only if $c = 1, \cdots$ 
The first  equation in Equation {\bf \ref{two.zero.conditions.21}} admits a solution  
if $\ws \not \in V^{\, 1}$.

\subsubsection{From two zero-conditions to one non-zero-condition} 
\label{translating.02}
Following paragraphs {\bf \ref{from.2.to.1}} and {\bf \ref{1.not.to.vanish}}, the two 
zero-conditions in Equation {\bf \ref{two.zero.conditions.21}} are equivalent to one non-zero-condition, 

\begin{equation}
\label{condition.21.04}
 V^{\, 1}_{\rrho} - V^{\, 2}_{\rrho - 1 - s_a + c \, \pp} \geq 1 + r_a - c \, p 
\end{equation}

\subsubsection{The stronger condition} 
\label{the.stronger.condition.02}
Equation ({\bf \ref{condition.21.04}}) is the statement that to eliminate the zeros, 
we want 
$V^{\, 1}_{\rrho} - V^{\, 2}_{\rrho - 1 - s_a + c \, \pp} \geq 1 + r_a - c \, p$, 
where $c = \{1, 2, \cdots\}$.
Since the row-lengths of a partition are by definition weakly decreasing, and 
$c = \{1, 2, \cdots\}$, this is the case if 
$V^{\, 1}_{\rrho} - V^{2, \rrho + \pp - s_a - 1}      \geq 1 + r_a - c \, p$.
In turn, is the case if 
$V^{\, 1}_{\rrho} - V^{\, 2}_{\rrho + \pp - s_a - 1}      \geq 1 + r_a -      p$. 
Thus, we should set $c=1$, to obtain 
\begin{equation}
\label{Burge.condition.01}
V^{\, 1}_ {\rrho} - V^{\, 2}_{\rrho + [\pp - s_a] - 1} \geq 1 - \ll p - r_a \rr
\end{equation}

\subsection{The third product} 
\label{den.conjugate.12}

\begin{equation}
\prod_{\bs \in W^{\, 1}} \ll  r_b \, \apos + s_b \, \aneg  + A^{  }_{\bs, W^{\, 2}} \, \apos 
                                                           - L^{++}_{\bs, W^{\, 1}} \, \aneg \rr
\end{equation}

\noindent vanishes if any factor satisfies

\begin{equation}
\label{zeros.conjugate.12}
\ll r_b + A^{  }_{\bs, W^{\, 2}} \rr \, \apos 
+
\ll s_b - L^{++}_{\bs, W^{\, 1}} \rr \, \aneg = 0
\end{equation}

\noindent which leads to the conditions

\begin{equation}
\label{conditions.conjugate.12.01}
-  A_{\bs, W^{\, 2}}   =       r_b + c \,   p,  
\quad
   L_{\bs, W^{\, 1}}   = - 1 + s_b + c \, \pp,
\end{equation}

\subsubsection{The stronger condition} Using the same arguments as in subsections {\bf \ref{den.12}} 
and {\bf \ref{den.21}}, are possible for $c = 0, 1, \dots$, $\bs \in W^{\, 1}$, 
$\bs \not \in W^{\, 2}$, and we should choose $c = 0$ to obtain  

\begin{equation}
\label{Burge.condition.02}
W^{\, 2}_{\rrho} - W^{\, 1}_{\rrho + s - 1} \geq 1 - r 
\end{equation}

\subsection{The fourth product} 
\label{den.conjugate.21}

\begin{equation}
\prod_{\bs \in W^{\, 2}} \ll r_b \, \apos + s_b \, \aneg - A^{++}_{\bs, W^{\, 1}} \, \apos 
                                                         + L^{  }_{\bs, W^{\, 2}} \, \aneg \rr
\end{equation}

\noindent vanishes if any factor satisfies  

\begin{equation}
\label{zeros.conjugate.21}
  \ll r_b - A^{++}_{\bs, W^{\, 1}} \rr \, \apos 
+ \ll s_b + L^{  }_{\bs, W^{\, 2}} \rr \, \aneg  = 0,
\end{equation}

\noindent which leads to the conditions

\begin{equation}
\label{conditions.conjugate.21.01}
- A^{ }_{\bs, W^{\, 1}} = 1 - r_b + c\,   p, 
\quad
  L^{ }_{\bs, W^{\, 2}} =   - s_b + c\, \pp 
\end{equation}

\subsubsection{The stronger condition} Using the same arguments as in subsections {\bf \ref{den.12}} 
and {\bf \ref{den.21}}, are possible for $c = 1, 2, \cdots$, $\square \in W^{\, 2}$ and $\square \not \in W^{\, 1}$, 
and we should choose $c=0$ to obtain

\begin{equation}
\label{weak.Burge.condition.01}
W^{\, 1}_{\rrho} - W^{\, 2}_{\rrho + [\pp - s_b]} \geq - \ll p - r_b \rr
\end{equation}

\subsection{The Burge conditions}

Equations {\bf \ref{weak.Burge.condition.02}} and {\bf \ref{Burge.condition.01}} are 
conditions on the partition pair {\bf V} on one side of $\cZ_{building.block}$, while 
Equations {\bf \ref{Burge.condition.02}} and {\bf \ref{weak.Burge.condition.01}} are 
conditions on the partition pair {\bf W} on the other side of $\cZ_{building.block}$.
When copies of $\cZ_{building.block}$ are glued to form conformal blocks, partition pairs 
on one side are identified with partition pairs on the other side. Thus each partition 
pair must satisfy all conditions. However, these conditions are not independent as two 
of them are satisfied when the other two are satisfied. More specifically, following 
paragraphs {\bf \ref{from.2.to.1}} and {\bf \ref{1.not.to.vanish}}, we can see that it 
is sufficient to enforce the two conditions in Equations {\bf \ref{Burge.condition.01}} 
and {\bf \ref{Burge.condition.02}}, 

\begin{equation}
\label{Burge.conditions}
\boxed{
\mu_{\rrho} - \mu_{\rrho + [\pp - s] - 1} \geq 1 - \ll p - r \rr, \quad
\nu_{\rrho} - \nu_{\rrho +        s  - 1} \geq 1         - r
}
\end{equation}

\noindent which we write in terms of a partition pair $\{\mu, \nu\}$ that could be 
on either side of $\cZ_{building.block}$. Partition pairs that satisfy the conditions 
in Equation {\bf \ref{Burge.conditions}} first appeared in the work of W H Burge on 
Rogers-Ramanujan-type identities \cite{burge}. The appeared in earlier studies of Virasoro 
characters in \cite{foda.lee.welsh, foda.welsh, welsh} and more recently in the context 
of the AGT correspondence in \cite{alkalaev.belavin, bershtein.foda}.

\section{Comments and open questions}
\label{section.11.comments}

\subsubsection*{Outline of result} We can generate conformal blocks of Virasoro $A$-series 
minimal models, labelled by the co-primes $p$ and $\pp$, times a Heisenberg factor, as follows. 
{\bf 1.} Start from the refined topological vertex of \cite{ikv} defined in Equation 
{\bf \ref{refined.topological.vertex}}, 
{\bf 2.} Glue four copies of the refined topological vertex to produce a 5D $U(2)$ basic web 
partition function, then take the $R \rightarrow 0$ limit, to obtain its 4D counterpart 
$\cW_{\bf V W \Delta}^{\, 4D}$ as in Equations {\bf \ref{W.5D.norm}, \ref{w.numerator.5D}, \ref{w.denominator.5D}}. 
{\bf 3.} Set the K{\"a}hler parameters ${\bf \Delta}$, and the deformation parameters $q$ and $t$ in 
$\cW_{\bf V W \Delta}^{\, 4D}$ as in Equations {\bf \ref{identification.12}}, 
                                               {\bf \ref{minimal.model.charges}},
					       {\bf \ref{neg.0.pos}}, and
					       {\bf \ref{parameters.03}},
and
{\bf 4.} require each of the partition pairs {\bf V} and {\bf W} to satisfy the Burge conditions
in Equation {\bf \ref{Burge.conditions}}.

\subsubsection*{The refined topological vertex of Awata and Kanno} We have used the refined topological 
vertex of Iqbal, Kozcaz and Vafa \cite{ikv}, but could have equally well used that of Awata and Kanno 
\cite{awata.kanno.01, awata.kanno.02}. The two vertices are equivalent as explained in 
\cite{awata.feigin.shiraishi}. 

\subsubsection*{Layers} The topic discussed in this note is vast and consists of many layers. 
We could have started our discussion from M-theory and used the language of M5 branes, but we 
decided to stay away from this, in this short note. 
Instead, we started from A-model topological strings, which live in a corner of the M-theory. 
From the refined topological vertex and topological strings, we obtained the building block of 
the instanton partition function of a 5D quiver gauge theory. We could have used the K-theoretic 
version of the AGT correspondence to obtain the minimal model analogues of the $q$-deformed 
Liouville conformal blocks discussed in \cite{awata.yamada.01, awata.yamada.02, aganagic.01, 
aganagic.02, nieri.01, nieri.02}. 
Instead, we skipped the $q$-deformed blocks, took the 4D limit, and used the 4D version of AGT 
to obtain minimal model conformal blocks. The M-theoretic origins of the minimal conformal blocks 
and their $q$-deformations should be topics of separate studies.

\subsubsection*{Interpretation} Missing from this note is an interpretation of the minimal model 
parameters in topological string or gauge theory terms. A complete interpretation will require 
working in M-theory terms which lies outside the scope of this work. In particular, missing is
an interpretation of the Burge conditions in topological string or gauge theory terms. 
We conjecture that such interpretations require re-derivations of existing results in ways that 
allow {\it ab initio} for minimal model parameters. Current derivations do not do that, and for
that reason, one obtains results that are not well-defined upon substitution of minimal-model 
parameters and that require {\it aposteriori} restrictions as in 
\cite{alkalaev.belavin, bershtein.foda} 
\footnote{\,
OF wishes to thank J-E Bourgine for discussions on this point.
}. We plan to address this topic in future work.

\subsubsection*{Postscript} Following the completion of work on this note, we became aware of 
the fact that $U(N)$ versions of Equations {\bf 2.14} and {\bf 2.17} in this note, were obtained 
in Equation {\bf 4.67} in \cite{bao.01}, and in Equation {\bf 5.1} and subsequent equations in 
\cite{mitev.03}
\footnote{\, 
The {\it basic web} in this note is related, by {\it \lq a flop\rq\,} to the {\it a strip} in 
\cite{bao.01}. This term was introduced in \cite{iqbal.kashani.poor.01}, where strip partition 
functions were studied in the context of the original, unrefined topological vertex.
}. 

\section*{Acknowledgements}
OF would like to thank M Bershtein for collaboration on \cite{bershtein.foda}, results of which 
were used in this note, and H Awata, J-E Bourgine, D Krefl, V Mitev, E Pomoni, A Tanzini and Y 
Zenkevich for discussions and useful remarks at various stages of work on this note. We thank 
the anonymous referee for many useful comments that helped us improve the presentationh. OF is 
supported by the Australian Research Council [ARC].


\end{document}